\documentclass[aps,prd,twocolumn,showpacs,nofootinbib,eqsecnum,
superscriptaddress]{revtex4-1}

\usepackage[centertags]{amsmath}
\usepackage{amssymb}
\usepackage{latexsym}
\usepackage{enumerate}
\usepackage{graphicx}
\usepackage{mathrsfs}
\usepackage{hyperref}
\usepackage{stmaryrd}
\usepackage{import}
\usepackage{tensor}
\usepackage{color}
\usepackage{bm}
\usepackage{multirow}
\usepackage{breakurl}
\usepackage{float}
\usepackage{lipsum}
\usepackage{siunitx}
\usepackage{comment}
\usepackage{mathtools}

\usepackage{stmaryrd}
\usepackage{color}
\usepackage{amsmath}
\usepackage{amssymb}
\usepackage{latexsym}
\usepackage{enumerate}
\usepackage{graphicx}
\usepackage{mathrsfs}
\usepackage{physics}
\usepackage{tabularx}
\makeindex

\newcommand{\be}{\begin{equation}}
\newcommand{\ee}{\end{equation}}
\newcommand{\bea}{\begin{eqnarray}}
\newcommand{\eea}{\end{eqnarray}}
\newcommand{\ba}{\begin{array}}
\newcommand{\ea}{\end{array}}
\newcommand{\bi}{\begin{itemize}}
\newcommand{\ei}{\end{itemize}}

\renewcommand{\l}{\left(}
\renewcommand{\r}{\right)}
\renewcommand{\a}{\alpha}
\renewcommand{\b}{\beta}
\newcommand{\g}{\gamma}
\newcommand{\G}{\Gamma}
\renewcommand{\d}{\delta}
\newcommand{\D}{\Delta}
\newcommand{\eps}{\epsilon}
\newcommand{\La}{\Lambda}
\newcommand{\la}{\lambda}
\newcommand{\vp}{\varphi}
\renewcommand{\O}{\Omega}
\renewcommand{\o}{\omega}
\renewcommand{\th}{\theta}
\newcommand{\Th}{\Theta}
\newcommand{\q}{\quad}
\renewcommand{\qq}{\qquad}
\newcommand{\s}{\sigma}
\newcommand{\Sig}{\Sigma}

\newcommand{\pa}{\partial}

\newcommand{\lhat}{\hat{l}}

\begin{document}

\title{Repeated faint quasinormal bursts in extreme-mass-ratio-inspiral
waveforms: Evidence from 
frequency-domain scalar self-force calculations on generic Kerr orbits}
\author{Zachary Nasipak}
\affiliation{Department of Physics and Astronomy, University of North 
Carolina, Chapel Hill, North Carolina 27599, USA}
\author{Thomas Osburn}
\affiliation{Department of Physics and Astronomy, State University of New 
York at Geneseo, New York 14454, USA}
\affiliation{Department of Physics and Astronomy, University of North 
Carolina, Chapel Hill, North Carolina 27599, USA}
\author{Charles R. Evans}
\affiliation{Department of Physics and Astronomy, University of North 
Carolina, Chapel Hill, North Carolina 27599, USA}

\begin{abstract}
We report development of a code to calculate the scalar self-force 
on a scalar-charged particle moving on generic bound orbits in the Kerr 
spacetime.  The scalar self-force model allows rapid development of 
computational techniques relevant to generic gravitational extreme-mass-ratio 
inspirals (EMRIs).  Our frequency-domain calculations are made with arbitrary 
numerical precision code written in \textit{Mathematica}.  We extend spectral 
source integration techniques to the Kerr spacetime, increasing computational 
efficiency.  We model orbits with nearly arbitrary inclinations 
$0\leq\iota<\pi/2$ and eccentricities up to $e \lesssim 0.8$.  This effort 
extends earlier work by Warburton and Barack where motion was restricted to 
the equatorial plane or to inclined spherical orbits.  Consistent with a 
recent discovery by Thornburg and Wardell (2017) in time-domain 
calculations, we observe self-force oscillations during the radially outbound 
portion of highly eccentric orbits around a rapidly rotating black hole.  As 
noted previously, these oscillations reflect coupling into the self-force by 
quasinormal modes excited during pericenter passage.  Our results confirm the 
effect with a frequency-domain code.  \emph{More importantly, we find that 
quasinormal bursts (QNBs) appear directly in the waveform following each 
periastron passage.}  These faint bursts are shown to be a superposition of 
the least-damped overtone (i.e., fundamental) of at least four ($l=m \le 4$) 
quasinormal modes.  Our results suggest that QNBs should appear in 
gravitational waveforms, and thus provide a gauge-invariant signal.  
Potentially observable in high signal-to-noise ratio EMRIs, QNBs would provide 
high-frequency components to the parameter estimation problem that would 
complement low-frequency elements of the waveform.
\end{abstract}


\maketitle

\section{Introduction}
\label{sec:intro}

Recent direct detections of gravitational waves have inaugurated a new branch
of multimessenger astronomy.  These observations of compact binary mergers by
advanced LIGO and advanced Virgo \cite{LVC1602.03839,LVC1606.04855,
LVC1706.01812,LVC1711.05578,LVC171016,LVC1709.09660} have led to discovery of
a new class of heavy stellar mass black holes, confirmed the primary site of
the r-process for creation of heavy elements, provided strong-field tests of
general relativity \cite{LVC1602.03841}, placed limits on the astrophysical
environments of compact binaries \cite{LVC1602.03846}, and made connection
with other parts of astronomy \cite{LVC1710.05833,LVC1710.05834}.  Detection
rates are poised to increase following recent sensitivity enhancements in
LIGO and Virgo, eventual completion of KAGRA \cite{KAGRA}, and development
of LIGO-India \cite{Unni13}.  Ground-based detectors will be complemented by
the LISA mission \cite{AmarETC13,AmarETC17,NASA11,ESA12} recently approved by
the European Space Agency, which will be sensitive to gravitational waves in a
lower frequency band ($10^{-4}-10^{-1}$ Hz).

A prime target for LISA will be extreme-mass-ratio inspirals (EMRIs)
consisting of a small compact object of mass $\mu \simeq 1-60 M_{\odot}$
(neutron star or black hole) in orbit about a supermassive black hole
($M\sim 10^5-10^7 M_{\odot}$) \cite{BerrETC19}.  With a small mass ratio
$\epsilon = \mu/M \simeq 10^{-7}-10^{-4}$, a gradual, adiabatic inspiral occurs,
which provides a natural application of black hole perturbation theory (BHPT)
and attendant gravitational self-force (GSF) calculations.  Once an EMRI
crosses into the detector passband, its orbital motion will accumulate a total
change in phase of order $\epsilon^{-1} \sim 10^{4}-10^{7}$ radians prior to
merger, with the implication that the small black hole will skim close to the
event horizon hundreds of thousands of times and provide an unprecedented
test of general relativity \cite{VigeHugh10,BaraCutl07,BrowETC07,AmarETC14,
BabaETC17}.  LISA will also serve as a cosmological probe, detecting EMRIs
out to redshifts of $z\sim 1-3$ \cite{AmarETC07,Bara09,AmarETC14,BabaETC17}.

Waveform templates produced from self-force calculations will be useful in
aiding signal detection of EMRIs and be essential for parameter estimation,
supplanting kludge waveforms derived from adiabatic inspiral calculations
\cite{BabaETC17}.  Long term self-force inspiral calculations of Schwarzschild
EMRIs are well advanced \cite{WarbETC12,OsbuWarbEvan16,WarbOsbuEvan17,
VandWarb18}, tracking the accumulated orbital or gravitational wave phase to
accuracies better than $\phi \simeq 0.1$ due to all
first-order-in-the-mass-ratio effects at post-1-adiabatic order
\cite{HindFlan08}, lacking only the orbit-averaged dissipative part of the
second-order self-force.  Progress is also being made on understanding and
calculating the second-order GSF \cite{Poun12a,Poun12b,PounMill14,WardWarb15,
MillWardPoun16,Poun17,MoxoFlan17}.  In the case of Kerr EMRIs, steady
developments have been made in GSF calculations for circular and bound
equatorial orbits \cite{ShahFrieKeid12,IsoyETC14,VandShah15,Vand16,MerlETC16,
FujiETC17,BaraGiud17}.  Progress has now been reported \cite{Vand17} in
calculating the GSF on generic Kerr orbits.  In principle this latest
self-force result could serve as the basis for long-term inspiral models of
astrophysically relevant EMRIs, but prospects are dimmed at present by high
computational costs of these GSF calculations.

In the past, the scalar field self-force analogue \cite{Quin00}, where a
scalar point charge orbiting a black hole sources a scalar wave that acts
back on the charge, has frequently been used as a simplified model to provide
understanding and to develop tools for use in the gravitational case.  The
scalar self-force (SSF) has been computed in Schwarzschild spacetime
\cite{Burk00a,Wise00,Burko00b,BaraBurk00,DetwMessWhit03,DiazETC04,Haas07,
VegaDetw08,VegaETC09,CasaETC09,CaniSopu09,CaniSopuJara10,WardETC12,DienETC12,
VegaETC13,WardETC14} and in Kerr spacetime using frequency domain (FD)
\cite{WarbBara10,WarbBara11,Warb15} and time domain (TD)
calculations \cite{ThorWard17}.

The present work generalizes the previous FD SSF calculations of Warburton
and Barack to arbitrary eccentric inclined orbits in Kerr spacetime.  Part
of our procedure involves calculating modes with the Mano-Suzuki-Takasugi
(MST) analytic function expansion approach
\cite{ManoSuzuTaka96a,ManoSuzuTaka96b,SasaTago03} using \textit{Mathematica}.
The code serves as a test bed for developing more advanced physical and
numerical techniques to aid downstream work in making generic Kerr GSF
calculations more practical.  For example, in this paper we adapted spectral 
source integration (SSI) \cite{HoppETC15} to the Kerr generic-orbit source 
problem, significantly optimizing computational efficiency.  Physically, we 
are able to explore rapidly the SSF in interesting high eccentricity and high 
black hole spin systems and follow-up work will examine the behavior of 
resonant-orbit configurations.

A primary physical result in this paper is confirming with our FD calculations 
the existence of quasinormal mode excitations in the self-force, which was 
discovered by Thornburg in TD SSF simulations of highly eccentric Kerr 
orbits.  This finding was discussed in a series of talks 
\cite{ThorCapra14,ThorCapra16,ThorCapra17} by 
Thornburg and reported in a paper by Thornburg and Wardell 
\cite{ThorWard17}.  Oscillations 
are observed in the self-force during the outbound portion of certain highly 
eccentric orbits following periastron passage near a rapidly rotating black 
hole.  These oscillations were confirmed to fit the least-damped overtone of 
the $l=m=1$ quasinormal mode.  We see precisely the same behavior in the 
self-force in our FD calculations (see \ref{sec:eqOrbits}) of a similar
highly eccentric ($e=0.8$) equatorial orbit about a rapidly rotating 
($a/M = 0.99$) primary.

\begin{figure}[ht!]
\includegraphics[width=0.95\columnwidth]{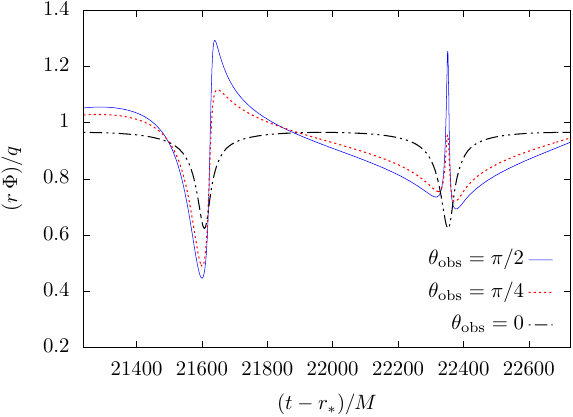}
\caption{The asymptotic waveform $r\Phi/q$ visible to observers at several 
polar angles: $\th_\text{obs}=\pi/2$ (blue solid line), 
$\th_\text{obs}=\pi/4$ (red dashed line), $\th_\text{obs}=0$ (black dot 
dashed line).  The plot window covers two radial librations.  Computed from 
an eccentric equatorial orbit (with associated apsidal advance), the waveform 
is biperiodic.  Sharp transitions roughly correspond to the retarded time of 
successive periastron passages.} 
\label{fig:waveform}
\end{figure}

\begin{figure}[ht!]
\includegraphics[width=0.95\columnwidth]{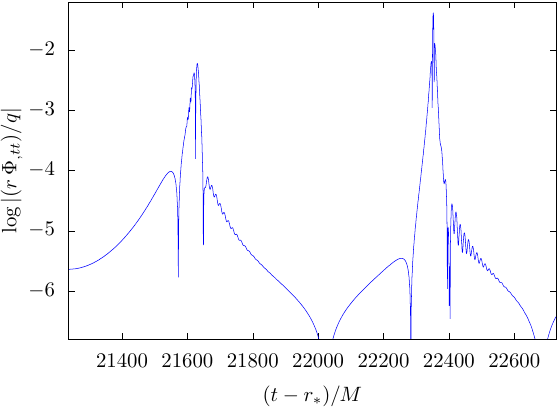}
	\caption{Log (base 10) of the absolute value of the second time 
derivative of the waveform in Fig.~\ref{fig:waveform} 
(for the observer at $\th_\text{obs}=\pi/2$).  The second time 
derivative enhances higher frequencies, making the faint QNBs visible in 
the aftermath of each periastron passage.} 
\label{fig:waveformDt2}
\end{figure}

More interestingly, we decided to take a look at the waveform in 
this same model to see if the excitation is imprinted in an 
asymptotically accessible signal.  Confirming our expectation, it is indeed 
possible to discern repeated (albeit faint) quasinormal bursts (QNBs) in the 
waveform following each periastron passage.  Fig.~\ref{fig:waveform} shows 
the asymptotic waveform over a period of two radial librations at several 
observer angles.  Without further processing, no quasinormal oscillations are 
directly apparent.  However, by high-pass filtering or otherwise enhancing 
high frequencies in the signal, we can make the low-level QNBs evident.  One 
particular way of enhancing high frequencies is shown in 
Fig.~\ref{fig:waveformDt2} where the log (base 10) of the 
absolute value of the second time derivative of the waveform is plotted.  
(Computing the second derivative is reminiscent of some numerical relativity 
codes where, to extract gravitational radiation, $\psi_4$ is first obtained, 
from which the waveforms are derived by integrating twice or by Fourier 
processing.)  Now the QNBs are revealed, superimposed on the lower frequency 
waveform components.  Use of a high-pass filter has similar effect (see 
\ref{sec:fittingQNM}).  We show in that later section that the QNBs are in 
fact a superposition of (at least) four least-damped quasinormal modes, with 
$l=m=1$ through $l=m=4$.

Our scalar self-force results suggest that comparable QNBs may appear 
in the gravitational waveform, which would provide a gauge-invariant signal 
of the effect.  These bursts are faint and might be fainter still in the 
gravitational case where $l=m=2$ will be the first mode excited.  On the 
other hand, we have not yet conducted a thorough parameter survey to find 
where the excitation is maximized.  Furthermore, it is entirely possible 
that even faint QNBs might be detected and measured using template matching.  
QNBs in EMRIs provide the exciting possibility of measuring black hole 
properties by repeatedly ``tickling the dragon's tail,'' as opposed to 
settling for the single final excitation of quasinormal modes seen in 
LIGO/Virgo mergers.  Finally, QNBs might reveal the presence of EMRIs in 
systems with heavy $M \gtrsim 10^7 M_\odot$ primaries, where the usual, 
low-frequency parts of the signal are difficult to detect but the periodic, 
higher-frequency QNBs lie in LISA's area of best sensitivity.\footnote{During 
discussion at a recent (May 2019) LISA Waveform Working Group meeting we were 
made aware that this signal has been observed in a gravitational self-force 
code;  M.~van de Meent, private communication.}  

The layout of this paper is as follows.  Section \ref{sec:formalism} covers
the formalism, with Sec.~\ref{sec:ssf} discussing the general nature of the
scalar self-force, Sec.~\ref{sec:orbit} reviewing the generic geodesic
motion problem and setting our notation, and Sec.~\ref{sec:teuk} outlining the
Fourier-harmonic decomposition of the scalar Teukolsky equation.  Section
\ref{sec:techniques} gives key details about the techniques we developed
and adapted to efficiently handle each phase of the generic Kerr SSF problem
including spectral solution of the orbital motion (Sec.~\ref{sec:orbitSSI}),
optimizations of the MST method for solving the homogeneous wave equation
(Sec.~\ref{sec:MST}), and efficient spectral source integration for solutions
to the inhomogeneous wave equation (Sec.~\ref{sec:kerrSSI}).
Section \ref{sec:regularize} discusses the regularization procedures and
computation of all four components of the scalar self-force.  We also discuss
there the split between conservative and dissipative parts of the self-force
on generic, nonresonant Kerr orbits.  In Sec.~\ref{sec:results} we present
our results, including the QNB-in-waveform discovery highlighted above, and 
discuss various implications.  For this paper we use units such that $c=G=1$, use metric 
signature $\l - + + + \r$, and the sign conventions of Misner, Thorne, and
Wheeler \cite{MisnThorWhee73}.

\section{Review of the Formalism for the Scalar Self-Force Problem}
\label{sec:formalism}

\subsection{The scalar self-force}
\label{sec:ssf}

The SSF model we consider assumes a point particle of mass
$\mu$ and scalar-charge $q$ in bound motion about a Kerr black hole of mass
$M$ and spin parameter $a$.  Perturbations in the gravitational field and the
associated GSF are neglected.  Instead the particle motion generates a scalar
field $\Phi$, whose local behavior acts back on the scalar charge to produce
the SSF.  Absent the SSF, the motion of the particle is a geodesic in the Kerr
spacetime.  The scalar field satisfies the curved-space Klein-Gordon equation
(i.e., the spin-0 Teukolsky equation \cite{Teuk73})
\be
\label{eqn:KGintro}
g^{\a\b}\nabla_\a\nabla_\b \Phi = -4\pi \rho,
\ee
where $\rho$ is the scalar (point) charge density and $g^{\a\b}$ is the
(inverse) Kerr metric.  Causal boundary conditions are selected, making the
resulting solution the retarded field $\Phi^\text{ret}$.  The particle's
timelike worldline is $x^\a_p(\tau)$ and its four-velocity is
$u^\a = d x_p^\a/d\tau$, where $\tau$ is proper time.  Formally, the SSF will
make the motion nongeodesic and the SSF will in principle depend upon the
entire past inspiral.  However, if $q$ is sufficiently small and the SSF weak,
the inspiral will be adiabatic, mimicking the GSF case with EMRIs.  Making this
assumption here, we take the past worldline as some (arbitrary) bound geodesic
and calculate the SSF along that fixed motion, the result being the
(approximate) \emph{geodesic self-force}.  While not a topic of this paper,
once the geodesic SSF is obtained in this way, it might be used in an
osculating elements calculation to determine the inspiral as is done with the
GSF \cite{PounPois08b,GairETC11,WarbETC12,OsbuWarbEvan16,WarbOsbuEvan17}.
The multiple periodicity of the background geodesic makes it possible to
solve the field equation in the FD, which we do in this paper.

The retarded field diverges at the point charge, necessitating a
regularization procedure \cite{Quin00} to compute the SSF.  Detweiler and
Whiting \cite{DetwWhit03} gave one particular separation of the retarded
field into regular and singular pieces
$\Phi^\text{ret} = \Phi^\text{R} + \Phi^\text{S}$, where $\Phi^\text{S}$
satisfies the same inhomogeneous wave equation \eqref{eqn:KGintro} as
$\Phi^\text{ret}$ but with (different) boundary conditions such that
$\Phi^\text{R}$ not only satisfies the source-free wave equation but is the
part of the field solely responsible for the SSF
\begin{align}
\label{eqn:SSF}
u^\b\nabla_\b(\mu u_\a) =
F_\a = \lim_{x \rightarrow x_p}q \nabla_\a \Phi^\text{R} .
\end{align}
Because the SSF is not orthogonal to the four-velocity \cite{Quin00}, all four
components of $F_\a$ must be determined.  Substitution of $\Phi^\text{ret}$ or
$\Phi^\text{S}$ in \eqref{eqn:SSF} in place of $\Phi^\text{R}$ produces
corresponding forces, $F^\text{ret}_\a$ and $F^\text{S}_\a$, both of which are
divergent on the particle worldline.  Thus even though one might write
\begin{align}
\label{eqn:ssfDW}
F_\a = F^\text{ret}_\a - F^\text{S}_\a ,
\end{align}
the expression is not immediately useful given the divergences.  Instead, one
practical procedure is mode-sum regularization \cite{BaraOri00,BaraOri03a},
wherein the retarded, singular, and regular fields (as well as their associated
forces) are decomposed into angular harmonics (typically using scalar
spherical harmonics $Y_{lm}$ for everything including components of vectors).
The individual mode amplitudes are finite and if the subtraction in
\eqref{eqn:ssfDW} is taken before summing (over $l$), the finite SSF is
recovered
\begin{align}
\label{eqn:ssfModeSum}
F_\a = \sum_{l=0}^{+\infty}
\left( F^{\text{ret},l}_{\a} - F^{\text{S},l}_{\a} \right) .
\end{align}
The singular part, $F^{\text{S},l}_{\a}$, can be obtained by local analytic
expansion in an $l$-dependent series with $l$-independent regularization
parameters. The lower-order parameters are known \cite{BaraOri03a}.  The
structure of higher-order terms is also understood \cite{DetwMessWhit03} and
analytic expressions have been given for certain restricted motions on
Schwarzschild \cite{HeffOtteWard12a,HeffETC17} and Kerr \cite{HeffOtteWard14}
backgrounds.  We fit numerically \cite{DetwMessWhit03} for higher-order
parameters in our more general application (Sec.~\ref{sec:regularize}).

With an assumed fixed background geodesic, the SSF can be decomposed
\cite{WarbBara11} into dissipative ($F^\text{diss}_\a$) and conservative
($F^\text{cons}_\a$) pieces
\begin{align}
F_\a = F^\text{diss}_\a + F^\text{cons}_\a ,
\end{align}
though assembling these pieces of the SSF from symmetries of the retarded
field is more subtle for generic, nonresonant orbits on Kerr.  The
dissipative part $F^\text{diss}_\a$ is responsible for the secular orbital
decay producing the inspiral, while $F^\text{cons}_\a$ serves to perturb the
orbital parameters.  The dissipative self-force does not require
regularization, as it is derived from the difference between retarded and
advanced fields,
$\Phi^\text{diss} = \tfrac{1}{2}(\Phi^\text{ret}-\Phi^\text{adv})$.  The
regularization procedure is necessary to determine $F^\text{cons}_\a$.  This
decomposition is further discussed in Sec.~\ref{sec:consDispSSF}.

\subsection{Bound geodesic motion in Kerr spacetime}
\label{sec:orbit}

We briefly review the generic geodesic motion problem in Kerr spacetime to
set our notation for use later in the paper.  In Boyer-Lindquist coordinates
$(t,r,\th,\vp)$ a Kerr black hole of mass $M$ and spin $a$ has the line element
\begin{multline}
\label{eqn:kerrLineElement}
ds^2 = -\l 1-\frac{2Mr}{\Sig}\r dt^2 +\frac{\Sig}{\D}dr^2
- \frac{4Mar\sin^2\th}{\Sig}dtd\vp \\
+\Sig d\th^2+\frac{\sin^2\th}{\Sig}\l \varpi^4-a^2\D \sin^2\th \r d\vp^2 ,
\end{multline}
where
\begin{align}
\Sig &=r^2+a^2\cos^2\th , \\
\D &=r^2-2Mr+a^2 , \\
\varpi &= \sqrt{r^2 + a^2} .
\end{align}
We define the conserved specific energy and $z$-component of the specific
angular momentum
\begin{align}
\mathcal{E} &= -\xi^\mu_{(t)}u_\mu = -u_t , \\
\mathcal{L}_z &= \xi^\mu_{(\vp)}u_\mu=u_\vp ,
\end{align}
using the Killing vectors $\xi^\mu_{(t)}$ and $\xi^\mu_{(\vp)}$, and define
the (scaled) Carter constant
\be
Q = K^{\mu\nu}u_\mu u_\nu - (\mathcal{L}_z-a\mathcal{E})^2 ,
\ee
associated with the Killing tensor $K^{\mu\nu}$ \cite{WalkPenr70}.

The geodesic equations are then
\cite{Cart68,MisnThorWhee73,DrasFlanHugh05}
\begin{align}
\label{eqn:R}
\left(\Sig_p \frac{dr_p}{d\tau} \right)^2
&= \left[\mathcal{E} \varpi_p^2 - a\mathcal{L}_z \right]^2
-\D _p\left[r_p^2+(\mathcal{L}_z-a \mathcal{E} )^2+Q\right] \notag \\
&\equiv V_r(r_p) , \\
\label{eqn:Th}
\left(\Sig_p \frac{d\th_p}{d\tau}\right)^2
&= Q-\mathcal{L}_z^2 \cot^2 \th_p-a^2(1-\mathcal{E} ^2) \cos^2\th_p \notag \\
&\equiv V_\theta(\th_p) , \\
\label{eqn:Phi}
\Sig _p \frac{d\vp_p}{d\tau} &= \Psi^{(r)}(r_p)+\Psi^{(\th)}(\th_p) , \\
\label{eqn:T}
\Sig _p \frac{dt_p}{d\tau} &= T^{(r)}(r_p)+T^{(\th)}(\th_p) ,
\end{align}
where the separate $r$-dependent and $\theta$-dependent functions appearing
in the last two equations are
\begin{align}
\label{eqn:PhiR}
&\Psi^{(r)}(r) = a \mathcal{E} \left(\frac{\varpi^2}{\D}-1\right)
-\frac{a^2\mathcal{L}_z}{\D} , \\
\label{eqn:PhiTh}
&\Psi^{(\th)}(\th) = \mathcal{L}_z \csc^2\th , \\
\label{eqn:TR}
&T^{(r)}(r) = \mathcal{E} \frac{\varpi^4}{\D}
+ a \mathcal{L}_z\left(1-\frac{\varpi^2}{\D}\right) , \\
\label{eqn:TTh}
&T^{(\th)}(\th) = -a^2 \mathcal{E}\sin^2\th .
\end{align}

Instead of parametrizing the orbit by $\mathcal{E}$, $\mathcal{L}_z$, and
$Q$, alternative constants of the motion can directly characterize the size,
shape, and orientation of the orbit.  The potential in \eqref{eqn:R} is a
quartic polynomial in $r$ and has four roots, which we denote by the following
ordering: $r_1\ge r_2\ge r_3 \ge r_4$.  For a bound, stable orbit the two
largest roots are finite and give the limits of radial motion.  Analogous
to Keplerian orbits, these extrema serve to define an eccentricity $e$ and
a semi-latus rectum $p$
\be
r_{\text{max}} =r_1 \equiv \frac{pM}{1-e} ,
\quad \quad
r_{\text{min}} = r_2 \equiv \frac{pM}{1+e} .
\ee
Having fixed $e$ in this way, it is useful to follow \cite{DrasFlanHugh05}
in defining the dimensionless quantities $p_3$ and $p_4$ from the final two
roots for use in later expressions
\be
p_3 \equiv r_3 (1-e)/M ,
\quad \quad
p_4 \equiv r_4 (1+e)/M .
\ee
An inclination angle $\iota$ is then defined from $\mathcal{L}_z$ and
$Q$ \cite{Hugh01}
\be
\cos\iota\equiv \frac{\mathcal{L}_z}{\sqrt{\mathcal{L}_z^2+Q}} .
\ee
It is straightforward to choose $e$, $p$, and $\iota$ as orbital parameters
and then solve for $\mathcal{E}$, $\mathcal{L}_z$, and $Q$
\cite{Schm02,DrasHugh05}.

A key additional reparameterization is to switch from $\tau$ to Mino time
$\la$ \cite{Mino03}
\be
d\la = \Sig^{-1}_p d\tau ,
\ee
which allows the $r$ and $\th$ motions to separate
\be
\label{eqn:laeqns}
\left(\frac{dr_p}{d\la}\right)^2 = V_r(r_p) ,
\q\q \left(\frac{d\th_p}{d\la}\right)^2 = V_\theta(\th_p) .
\ee
These equations yield solutions that are functions of the new curve parameter
$\la$, e.g., $r_p(\la)$, with confusion over the slight abuse of notation
avoided by explicit reference to the new curve parameter.  The subscript $p$
continues to mean ``on the worldline.''  With this in mind, further
reparameterizations are made by introducing Darwin-like \cite{Darw61} angular
coordinates $\psi$ and $\chi$ \cite{Schm02,DrasHugh05}
\begin{align}
\label{eqn:newpars}
&r_p(\psi) = \frac{pM}{1+e\cos\psi} , \q\q
\cos\th_p(\chi)= \sqrt{z_-}\cos{\chi} , \\
&z_{\pm}\equiv \frac{\mathcal{L}_z^2+Q+\b
\pm\sqrt{(\mathcal{L}_z^2+Q+\b )^2-4Q\b }}{2\b } ,
\end{align}
where $\b \equiv a^2(1-\mathcal{E} ^2)$.  In the last equation,
$z_{\pm}$ are roots of $V_\theta$, with ordering of roots taken to be
$0 \le z_{-} \le 1 \le z_{+}$ and $z_-$ associated with the turning points.

Equations \eqref{eqn:laeqns} and \eqref{eqn:newpars} may be combined to find
differential equations relating $\psi$ and $\chi$ to $\la$, or vice versa with
functions $\la = \la^{(r)}(\psi)$ and $\la = \la^{(\theta)}(\chi)$ satisfying
\begin{align}
\label{eqn:Pr}
\frac{d\la^{(r)}}{d\psi}
&= \frac{a(1-e^2) \left[(p-p_4)+e(p-p_4\cos\psi)\right]^{-1/2}}{M\b ^{1/2}
\left[(p-p_3)-e(p+p_3\cos\psi)\right]^{1/2}} \notag \\
&\equiv P^{(r)}(\psi) ,
\\
\label{eqn:Pth}
\frac{d\la^{(\th)}}{d\chi}
&= \left[ \b (z_+ -z_-\cos^2 \chi ) \right]^{-1/2} \equiv P^{(\th)}(\chi) .
\end{align}
The definitions of $\psi$ and $\chi$ in \eqref{eqn:newpars} are made to
improve the behavior of the differential equations at what would otherwise be
turning points for $r$ and $\theta$.  The solutions for $\la^{(r)}$ and
$\la^{(\th)}$ can be expressed as integrals
\begin{align}
\label{eqn:lambdar}
&\la = \la^{(r)}(\psi) = \int_0^\psi P^{(r)}(\psi')\, d\psi'
+ \la_0^{(r)} ,
\\
\label{eqn:lambdath}
&\la = \la^{(\th)}(\chi) = \int_0^\chi P^{(\th)}(\chi')\, d\chi'
+ \la_0^{(\th)} ,
\end{align}
where $\la_0^{(r)}$ and $\la_0^{(\th)}$ are integration constants, with
$\la_0^{(r)}-\la_0^{(\th)} \ne 0$ providing initial conditions for orbits that
do not simultaneously pass through $r=r_{\text{min}}$ and
$\th=\th_{\text{max}}$.  The effect of choosing a nonzero value for
$\la_0^{(\th)}$, for example, is demonstrated in Fig.~\ref{fig:torus}.
The integrals in Eqs.~\eqref{eqn:lambdar} and \eqref{eqn:lambdath} may be
reexpressed in terms of elliptic integrals \cite{DrasHugh05,FujiHiki09} and
thereby regarded as solved.  We adopt an alternate approach in this paper,
based on results in \cite{HoppETC15} and the observation that the integrands in
Eqs.~\eqref{eqn:lambdar} and \eqref{eqn:lambdath} are smooth and periodic
functions.  This allows functions like $P^{(r)}(\psi)$ in Eq.~\eqref{eqn:Pr}
to be represented by exponentially convergent Fourier series that can be
accurately truncated at some $n=N-1$
\be
\label{eqn:Pcos}
P^{(r)}(\psi) \simeq \sum_{n=0}^{N-1}\tilde{\mathcal{P}}^{(r)}_n \cos(n\,\psi) .
\ee
Term-by-term integration of \eqref{eqn:lambdar} then gives
\begin{align}
\label{eqn:Pint}
\la^{(r)}(\psi) \simeq \tilde{\mathcal{P}}^{(r)}_0 \psi
+ \sum_{n=1}^{N-1} \frac{\tilde{\mathcal{P}}^{(r)}_n}{n} \sin(n\,\psi)
+ \la_0^{(r)} ,
\end{align}
with a similar expression for \eqref{eqn:lambdath}.  The Fourier series
coefficients ($\tilde{\mathcal{P}}^{(r)}_n$) are ostensibly derived
themselves from integrals, but it proves possible in a numerical calculation
to replace the Fourier series representation with the discrete Fourier
transform (DFT).  The coefficients in the DFT are then rapidly and
accurately obtained using the fast Fourier transform (FFT) algorithm.
Section \ref{sec:orbitSSI} details this new application of spectral integration
to Kerr orbits; reference \cite{HoppETC15} demonstrates the application to
integrating Schwarzschild geodesics.

The periods of motion in $r$ and $\th$ measured in Mino time are
\begin{align}
\La_r &= \la^{(r)}(2\pi) - \la_0^{(r)} , \q
\La_{\th} = \la^{(\th)}(2\pi) - \la_0^{(\th)} ,
\end{align}
and the corresponding frequencies with respect to Mino time are
\begin{align}
\Upsilon_r &= \frac{2\pi}{\La_r},
\q\q\q
\Upsilon_{\th} = \frac{2\pi}{\La_{\th}} .
\end{align}

Eqs.~\eqref{eqn:T} and \eqref{eqn:Phi} can be reexpressed in terms of Mino
time derivatives and the evolution of $t$ and $\vp$ in terms of $\la$ have
the following formal dependence
\begin{align}
\label{eqn:tpOfLa}
t_p(\la) &= \Gamma  \la + \D t^{(r)}(\la) + \D t^{(\th)}(\la) + t_0 ,
\\
\label{eqn:phipOfLa}
\vp_p(\la) &= \Upsilon_{\vp} \la + \D \vp^{(r)}(\la)
+ \D \vp^{(\th)}(\la) + \vp_0 ,
\end{align}
with $t_0$ and $\vp_0$ constants.  In these expressions the average rates of
accumulation of $t$ and $\vp$ in $\la$ are, respectively
\begin{align}
\Gamma &= \frac{1}{\La_r} \int_0^{\La_r} T^{(r)} \, d\la
+ \frac{1}{\La_\th} \int_0^{\La_\th} T^{(\th)} \, d\la ,
\\
\Upsilon_\vp &= \frac{1}{\La_r} \int_0^{\La_r} \Psi^{(r)} \, d\la
+ \frac{1}{\La_\th} \int_0^{\La_\th} \Psi^{(\th)} \, d\la ,
\end{align}
while $\D t^{(r)}$ and $\D \vp^{(r)}$ are oscillatory functions with period
$\La_r$ and $\D t^{(\th)}$ and $\D \vp^{(\th)}$ are oscillatory functions with
period $\La_\th$.  These oscillatory functions are described by similar
integrals, and we obtain their numerical solution via spectral integration
in like fashion to Eq.~\eqref{eqn:Pint} (see Sec.~\ref{sec:orbitSSI}).  The
average motion of $t$ and $\vp$, along with the Mino time frequencies, then
provide the fundamental (coordinate time) frequencies
\be
\O_r = \frac{\Upsilon_r}{\Gamma } , \q\q
\O_{\th} = \frac{\Upsilon_{\th}}{\Gamma } , \q\q
\O_{\vp} = \frac{\Upsilon_{\vp}}{\Gamma } .
\ee
The motion of the particle can then be described by a discrete frequency
spectrum
\be \label{eqn:freqMKN}
\o_{mkn} = m \O_\vp +k \O_\th + n \O_r ,
\ee
with $m$, $k$, and $n$ being integers.

\subsection{Scalar wave equation}
\label{sec:teuk}

The charge density $\rho$, which acts as the source of the wave equation
\eqref{eqn:KGintro}, is that of a point charge following the timelike
orbital motion
\begin{align}
\rho(t,r,\th,\vp) &= q \int \d^{(4)}(x^\a-x_p^\a(\tau)) \l -g \r ^{-1/2} d\tau, \\
&= q \frac{\d(r-r_p) \d(\cos\th-\cos\th_p)
\d(\vp-\vp_p)}{T^{(r)}(r_p)+T^{(\th)}(\th_p)}, \notag
\end{align}
where $\sqrt{-g} = \Sig \sin\th$ and $T^{(r)}$ and $T^{(\th)}$ are given by
Eqs.~\eqref{eqn:TR} and \eqref{eqn:TTh}, respectively.  The wave equation is
equivalent to the TD spin-0 Teukolsky equation \cite{Teuk73}
\begin{align}
\label{eqn:teuk0}
&\left(\frac{(r^2+a^2)^2}{\D}-a^2 \sin^2{\th} \right)\frac{\pa^2\Phi}{\pa t^2}
+\frac{4 M a\, r}{\D}\frac{\pa^2\Phi}{\pa t\pa \vp}
\notag \\
&\qquad\q + \left( \frac{a^2}{\D}-\frac{1}{\sin^2{\th}} \right)
\frac{\pa^2 \Phi}{\pa\vp^2}-\frac{\pa}{\pa r}\left(\D
\frac{\pa\Phi}{\pa r}  \right)
\\
&\qquad\qquad\q\q  - \frac{1}{\sin{\th}} \frac{\pa}{\pa\th}
\left( \sin{\th}\frac{\pa \Phi}{\pa \th}\right) = 4\pi \Sigma \rho .
\notag
\end{align}

\subsubsection{Separation of variables}

Equation \eqref{eqn:teuk0} is amenable to solution via separation of variables
\cite{BrilETC72,Teuk73}
\be
\label{eqn:fieldDecomp}
\Phi = \sum_{\hat{l}mkn} R_{\hat{l}mkn}(r)\, S_{\hat{l}mkn}(\theta)\,
e^{im\vp}\, e^{-i \o_{mkn} t} .
\ee
Here $R_{\hat{l}mkn}(r)$ is the Teukolsky radial function,
$S_{\hat{l}mkn}(\th)$ is the spheroidal Legendre function with $\hat{l}$ and
$m$ multipole indices and spheroidicity $\s^2=-a^2\o_{mkn}^2$ (hence the
$\hat{l}mkn$ subscripts).  In the above equation and henceforth, the following
condensed notion is introduced to represent the sum over modes
\begin{align}
\sum_{\hat{l}mkn} \equiv \sum_{\hat{l}=0}^{+\infty} \;
\sum_{m=-\hat{l}}^{\hat{l}} \, \sum_{k=-\infty}^{+\infty} \,
\sum_{n=-\infty}^{+\infty} .
\end{align}
Following Warburton and Barack \cite{WarbBara11}, we use $\hat{l}$ for the
spheroidal harmonic index and reserve $l$ for the spherical harmonic index used
in the mode-sum regularization.  The FD decomposition in
\eqref{eqn:fieldDecomp} assumes bound motion, with a resulting discrete
frequency spectrum that allows the field to be represented by a multiple
Fourier series.

We follow \cite{Hugh00,SasaNaka82,BrilETC72} in connecting the Teukolsky
function, $R_{\hat{l}mkn}(r)$, to a new radial function, $X_{\hat{l}mkn}(r)$
\begin{align}
\label{eqn:SNtrans}
X_{\hat{l}mkn}(r) = \sqrt{r^2 + a^2}\, R_{\hat{l}mkn}(r) .
\end{align}
(Warburton and Barack \cite{WarbBara10,WarbBara11,Warb15} make a different
transformation.)  Both $R_{\hat{l}mkn}$ and $X_{\hat{l}mkn}$ are used in
what follows (see especially Sec.~\ref{sec:techniques}).  Inserting
Eqs.~\eqref{eqn:fieldDecomp} and \eqref{eqn:SNtrans}
into Eq. \eqref{eqn:teuk0}, we arrive at two ordinary differential equations
for $X_{\hat{l}mkn}(r)$ and $S_{\hat{l}mkn}(\th)$
\begin{align}
\label{eqn:angTeuk}
&\bigg[\frac{1}{\sin\th}\frac{d}{d\th}\l\sin\th
\frac{d}{d\th}\r -\frac{m^2}{\sin^2\th} - a^2\o_{mkn}^2 \sin^2\th
\\
&\q\q\q\q\q - 2am\o_{mkn} - \la_{\lhat mkn} \bigg] S_{\lhat mkn}(\th) = 0,
\notag
\\
\label{eqn:genSN}
&\left[\frac{d^2}{dr_*^2} - U_{\hat{l}mkn}(r)\right] X_{\hat{l}mkn}(r)
= Z_{\hat{l}mkn}(r) ,
\end{align}
where $\la_{\lhat mkn}$ is the angular eigenvalue, or separation constant,
and $r_*$ is the tortoise coordinate
\be
r_{*} = r + \frac{2 M r_{+}}{r_{+}-r_{-}} \ln\frac{r-r_{+}}{2 M}
- \frac{2 M r_{-}}{r_{+}-r_{-}} \ln\frac{r-r_{-}}{2 M} ,
\ee
which follows from integrating
\begin{align}
\frac{dr_*}{dr} = \frac{\varpi^2}{\D} .
\end{align}
Here $r_{\pm} = M \pm \sqrt{M^2 - a^2}$ are the outer and inner horizon radii
(roots of $\D(r) = 0$).  Our definition of $r_{*}$ agrees with e.g.,
\cite{DrasFlanHugh05,SasaTago03} but differs from \cite{WarbBara10,
WarbBara11,Warb15}.  The radial potential $U_{\hat{l}mkn}(r)$ is
\begin{align}
&U_{\hat{l}mkn}(r) = \varpi^{-8} \Big[2am\,\o_{mkn} \, \varpi^6 -6 M a^4 r
-4 M a^2 r^3
\notag \\
&\q\q\q\q + a^2 \, \varpi^4 (1-m^2) + 8M^2 a^2 r^2 - \o_{mkn}^2 \,\varpi^8
\notag \\
&\q\q\q\q
+ \la_{\hat{l}mkn} \D \, \varpi^4 - 4M^2 r^4 +2M r^5 \Big],
\end{align}
and $Z_{\lhat mkn}(r)$ gives the radial behavior of the source in the FD
\be
\label{eqn:sourceDecomp}
\rho = -\frac{\varpi^3}{4 \pi \Sig \D} \sum_{\lhat mkn}
Z_{\lhat mkn}(r)\, S_{\lhat mkn}(\th)\, e^{im\vp}\, e^{-i\o_{mkn}t}.
\ee

\subsubsection{Radial solutions and time domain reconstruction}
\label{sec:radSol}

General solution of Eq.~\eqref{eqn:genSN} requires two independent
homogeneous solutions, $\hat{X}^{+}_{\lhat mkn}(r)$ and
$\hat{X}^{-}_{\lhat mkn}(r)$, that hold throughout the region
$r_{+}\le r \le \infty$ and have respective asymptotic dependence
\begin{align}
\label{eqn:Xp}
\hat{X}^+_{\lhat mkn}(r) &\simeq e^{ +i\o_{mkn}r_*}, \q\q r\rightarrow\infty ,
\\
\label{eqn:Xm}
\hat{X}^-_{\lhat mkn}(r) &\simeq e^{-i \gamma_{mkn}r_*} , \q\q
r \rightarrow r_{+} .
\end{align}
Here $\gamma_{mkn} = \o_{mkn}-m\o_+$ is the wave number at the horizon, with
$\o_+ = a/2Mr_+$ denoting the angular velocity of the event horizon.  The
solution $\hat{X}^+_{\lhat mkn}$ is ``outgoing,'' while the solution
$\hat{X}^-_{\lhat mkn}$ is ``downgoing.''  These two can be combined to
construct the causal Green function for the radial equation \eqref{eqn:genSN},
associated ultimately with the retarded solution in the TD.  The
solution of the inhomogeneous Eq.~\eqref{eqn:genSN} is then found to be
\begin{align}
\label{eqn:XsolGibbs}
&X_{\lhat mkn}^{\text{inh}} = c_{\lhat mkn}^+(r)\hat{X}_{\lhat mkn}^+(r)
+c_{\lhat mkn}^-(r)\hat{X}_{\lhat mkn}^-(r) ,
\\
\label{eqn:XsolGibbscPlus}
&c_{\lhat mkn}^+(r) = \int_{r_\text{min}}^r
\frac{\varpi(r')^2\hat{X}_{\lhat mkn}^-(r')Z_{\lhat mkn}(r')}
{W_{\lhat mkn} \D (r')} dr' ,
\\
\label{eqn:XsolGibbscMinus}
&c_{\lhat mkn}^-(r) = \int_r^{r_\text{max}}
\frac{\varpi(r')^2\hat{X}_{\lhat mkn}^+(r')Z_{\lhat mkn}(r')}
{W_{\lhat mkn} \D (r')} dr' ,
\end{align}
where
\be
W_{\lhat mkn} = \hat{X}_{\lhat mkn}^- \frac{d\hat{X}_{\lhat mkn}^+}{dr_*}
-\hat{X}_{\lhat mkn}^+ \frac{d\hat{X}_{\lhat mkn}^-}{dr_*},
\ee
is the (constant) Wronskian.

An attempt to use $X_{\lhat mkn}^{\text{inh}}(r)$ from \eqref{eqn:XsolGibbs}
with \eqref{eqn:SNtrans} in \eqref{eqn:fieldDecomp} to make a (time domain)
Fourier reconstruction of the field at points within the libration region
$r_\text{min} < r < r_\text{rmax}$ is fraught with difficulty due to Gibbs
oscillations caused by the delta function source.  In this region, at points
away from the worldline, the convergence in $k$ and $n$ is slow, while
derivatives (needed for the SSF) may not even converge at the particle.  The
usual path around
this problem, at least in a background spacetime with spherical symmetry, is
the method of extended homogeneous solutions (EHS) \cite{BaraOriSago08}.
In that case the four-dimensional wave equation separates into two-dimensional
wave equations in $t$ and $r$ for each spherical harmonic order $l$, $m$.
Extended homogeneous solutions are found mode by mode, which are finite at
the particle as needed for mode-sum regularization.  Unfortunately, in Kerr
spacetime the angular decomposition in spheroidal harmonics is inseparably
linked to the transformation into the FD.  As Warburton and Barack
\cite{WarbBara11} have shown however, it is still possible to define functions
on the spherical harmonic basis that can be extended to the particle location
and are finite there.

This procedure begins with determining normalization coefficients,
$C^\pm_{\lhat mkn}$, which are found by evaluating $c^\pm_{\lhat mkn}(r)$ at
the limits of the radial libration region
\be
\label{eqn:Z}
C^\pm_{\lhat mkn} = \int_{r_\text{min}}^{r_\text{max}}
\frac{\varpi^2\hat{X}^\mp_{\lhat mkn}(r)Z_{\lhat mkn}(r)}{W_{\lhat mkn}\D} dr ,
\ee
and which are used to define the properly normalized extended homogeneous
radial modes in the FD
\be
X^\pm_{\lhat mkn}(r) = C^\pm_{\lhat mkn}\hat{X}^\pm_{\lhat mkn}(r) .
\ee
These solutions in turn may be used in \eqref{eqn:fieldDecomp} to define
extended solutions in the full time and space domain
\be
\label{eqn:EHS1}
\Phi^\pm \equiv \frac{1}{\varpi} \sum_{\lhat mkn} X^\pm_{\lhat mkn}(r) \,
S _{\lhat mkn}(\th) \, e^{im\vp} \, e^{-i\o_{mkn}t} ,
\ee
from which the retarded solution to \eqref{eqn:teuk0}, at least off the
worldline, can be given as
\begin{align}
\label{eqn:EHS2}
\Phi^\text{ret}(t,r,\th,\vp) &=  \Phi^-(t,r,\th,\vp)\, \Th(r_p(t)-r)
\\
&\q\q\q + \Phi^+(t,r,\th,\vp)\, \Th(r-r_p(t)) . \notag
\end{align}
While the functions $\Phi^\pm$ \eqref{eqn:EHS1} converge exponentially in
$k$ and $n$ and their use eliminates the Gibbs behavior \emph{near} the
particle in the libration region, the full reconstruction \eqref{eqn:EHS2}
is not of immediate use in calculating the SSF.  The approach taken by
Warburton and Barack relies upon using the representation \cite{Hugh00b} of
spheroidal angular harmonics in terms of spherical harmonics
$Y_{lm}(\th,\vp)$
\be
\label{eqn:spheroidalSpherical}
S_{\lhat mkn}(\th)\,e^{im\vp}
= \sum_{l = |m|}^{+\infty} b^{\lhat kn}_{lm} \, Y_{lm}(\th,\vp) .
\ee
While the spheroidal harmonics of order $\lhat$ couple to an infinite number
of spherical harmonics, the coupling coefficients $b^{\lhat kn}_{lm}$ decay
in size rapidly as the difference in orders $|\lhat -l|$ grows
\cite{WarbBara10}, the rate dependent upon the spheroidicity $a^2 \o_{mkn}^2$.
In a numerical calculation, the number of spherical harmonics needed for a
given accuracy can be determined.  The coupling coefficients are determined
by a three-term recurrence relation that results from inserting
\eqref{eqn:spheroidalSpherical} into \eqref{eqn:angTeuk}.

Substituting \eqref{eqn:spheroidalSpherical} into \eqref{eqn:EHS1}, the
five-fold summation may be reordered to leave $l$ and $m$ for last.  This
allows the extended functions $\phi^\pm_{lm}(t,r)$ to be defined,
\begin{equation}
\label{eqn:phiLM}
\phi^\pm_{lm}(t,r) = \frac{1}{\varpi} \sum_{\lhat kn}
b^{\lhat kn}_{lm} \, X^\pm_{\hat{l}mkn}(r) \, e^{-i \o_{mkn} t} ,
\end{equation}
where in a practical numerical calculation the sum over $\lhat$ will be
finite in number, as will the sums over $k$ and $n$ given their exponential
convergence.  The remaining sums allow $\Phi^\pm$ to be recovered
\begin{equation}
\label{eqn:phiSpherical}
\Phi^\pm(t,r,\th,\vp) = \sum_{l = 0}^{+\infty} \sum_{m=-l}^{l}
\phi^\pm_{lm}(t,r) \, Y_{lm}(\th,\vp) .
\end{equation}
The functions $\phi^\pm_{lm}(t,r)$ are not modes in the fullest sense, since
there are no wave equations in $t$ and $r$ that they satisfy.  However, they
do derive from linear combinations of extended (homogeneous) radial modes
in the FD, they provide a decomposition of $\Phi^\pm$, and they are finite
at the location of the particle.  These properties are all that is essential
for employing mode-sum regularization, as shown by \cite{WarbBara10,WarbBara11,
Warb15} and as outlined in Sec.~\ref{sec:regularize}.  Our generalization
here to eccentric inclined orbits introduces no qualitatively new element in
the Kerr SSF regularization, only a further dimension in the mode calculations.

The homogeneous solutions $\hat{X}^\pm_{\hat{l}mkn}(r)$ are often
obtained by numerical integration \cite{WarbBara10,WarbBara11,Warb15} of
\eqref{eqn:genSN}.  In this work, however, we use a \textit{Mathematica} code
employing the MST method (Sec.~\ref{sec:MST}) to derive the mode functions.
The resulting code is very accurate but slow.  We are concurrently developing
and testing a faster, complementary C-code based on numerical integration
of \eqref{eqn:genSN}.

\section{Analytic and Numerical Solution Techniques}
\label{sec:techniques}

This section describes some of the analytic and numerical techniques we use
to solve the SSF problem in the generic Kerr case.  It provides some details
on spectral solution of the orbit equations, on efficient use of the MST
method to obtain certain mode functions, and especially on spectral
integration of source terms in the scalar case.  The computational roadmap
is as follows:

\begin{enumerate}[(i)]

\item After specifying the parameters $p$, $e$, $\iota$, and $a$, we solve
for the geodesic motion on Kerr using spectral integration techniques
(Sec.~\ref{sec:orbitSSI}).  From the geodesic, we determine the fundamental
frequencies of the orbit, $\O_r$, $\O_\th$, and $\O_\varphi$.

\item We calculate the radial and polar mode functions for each frequency
and multipole.  The polar mode functions (spheroidal harmonics) are constructed
using Eq.~\eqref{eqn:spheroidalSpherical}.  We calculate the homogeneous
radial mode functions, $\hat{X}^\pm_{\lhat mkn}$, using the MST function 
expansion formalism, with Sec.~\ref{sec:MST} serving primarily to discuss an 
efficient approach to finding the near-horizon modes.

\item Finally, we discuss in Sec.~\ref{sec:kerrSSI} means to evaluate the
normalization constants $C^\pm_{\lhat mkn}$, which determine the scalar
field via the EHS method, using spectral source integration techniques.  In
the scalar case, it proves possible to decompose the source integration
\eqref{eqn:Z} into products of one-dimensional integrals.

\end{enumerate}

\subsection{Spectral integration of the geodesic equations}
\label{sec:orbitSSI}

As an alternative to using initial value integration, or to using special
functions \cite{DrasHugh05,FujiHiki09}, we employ a spectral
(Fourier) integration technique to find the Kerr geodesics numerically.
Spectral integration of the orbital motion problem in Schwarzschild spacetime
was previously carefully laid out in \cite{HoppETC15}.  In this
subsection we generalize that approach to generic bound geodesics in Kerr
spacetime.

We first consider the dependence of $\la$ on Darwin angles $\psi$ and $\chi$.
The integration for $\la^{(r)}(\psi)$ is given as an example, but the same
approach applies to $\la^{(\th)}(\chi)$.  As discussed in
Sec.~\ref{sec:orbit}, the function $P^{(r)}(\psi)$ can be written as a cosine
series because it is smooth, even, and periodic
\begin{align}
\label{eqn:PcosFull}
P^{(r)}(\psi) &=\sum_{n=0}^{\infty} \tilde{\mathcal{P}}^{(r)}_n \cos(n\,\psi) .
\end{align}
Because $P^{(r)}$ is $C^{\infty}$, \eqref{eqn:PcosFull} converges exponentially
with the number of harmonics, and for a given accuracy may be truncated at
some $n=N_r - 1$ as in \eqref{eqn:Pcos}.

Fourier series coefficients like $\tilde{\mathcal{P}}^{(r)}_n$ are derived
from integrals, so computing many of these by, for example, adaptive stepsize
integration is no improvement over simply integrating \eqref{eqn:Pr} itself.
Instead, an efficient alternative is to use the discrete Fourier transform
(DFT).  To do so, we use \eqref{eqn:Pcos} to sample $P^{(r)}(\psi)$ at $N_r$
evenly spaced points $\psi_j$.  The $N_r$ sampled values $P^{(r)}(\psi_j)$ are
the DFT of $N_r$ Fourier coefficients, $\mathcal{P}^{(r)}_n$.  Up to a
normalization factor, the DFT coefficients (with no tilde) converge
exponentially to the Fourier series coefficients (with tilde) as the sample
number $N_r$ increases.  Since $P^{(r)}$ is even, we discretely sample the arc
of half the radial motion and represent the function with the type-I discrete
cosine transform (DCT-I)
\be
\psi_j \equiv \frac{j \pi}{N_r -1}, \q \q j \in {0, 1, \ldots, N_r -1},
\ee
\begin{align}
\label{eqn:dctPrti}
\mathcal{P}^{(r)}_n &= \sqrt{\frac{2}{N_r -1}}
\bigg[ \frac{1}{2} P^{(r)}(0) + \frac{1}{2} (-1)^n P^{(r)}(\pi) \notag \\
&\q\q\q\q\q\q +\sum_{j=1}^{N_r -2} P^{(r)}(\psi_j)
\cos \l n\psi_j \r \bigg] ,
\end{align}
\begin{align}
\label{eqn:dctPr}
P^{(r)}(\psi) &= \sqrt{\frac{2}{N_r -1}}
\bigg[ \frac{1}{2} \mathcal{P}^{(r)}_0 + \frac{1}{2}
\mathcal{P}^{(r)}_{N_r -1}\cos\left[ (N_r -1)\psi \right] \notag
\\
&\q\q\q\q\q\q +\sum_{n=1}^{N_r -2} \mathcal{P}^{(r)}_n
\cos \left(n\psi \right) \bigg] .
\end{align}
The DFT (or in this case DCT) may be computed numerically using a fast
Fourier transform (FFT) algorithm, efficiently finding all of the Fourier
coefficients $\mathcal{P}^{(r)}_n$.  The angular sampling of 
$P^{(\th)}(\chi)$ is made over $N_{\th}$ equally spaced points.  The 
required radial and angular sample numbers are independent and subject only 
to desired numerical accuracy goals.

Returning to the radial motion example, once $P^{(r)}(\psi)$ is adequately
represented, then $\la^{(r)}$ is found by substituting \eqref{eqn:dctPr} into
\eqref{eqn:lambdar} and integrating term-by-term
\begin{align}
\label{eqn:lambdapsi}
\la^{(r)}(\psi) &=
\sqrt{\frac{2}{N_r -1}} \bigg[ \frac{1}{2} \psi\,\mathcal{P}^{(r)}_0
+ \frac{1}{2} \mathcal{P}^{(r)}_{N_r -1}
\frac{\sin\left[ (N_r -1)\psi \right]}{(N_r -1)}
\notag \\
& \q\q\q\q\q\q +\sum_{n=1}^{N_r -2} \mathcal{P}^{(r)}_n
\frac{\sin \left(n\psi \right)}{n} \bigg] ,
\end{align}
an expression that can be evaluated at any $\psi$.  The same is done for
$\la^{(\th)}(\chi)$.  The Mino time periods, $\La_r$ and $\La_{\th}$, are
related to the leading Fourier coefficients
\begin{align}
\La_r &= \pi \mathcal{P}^{(r)}_0\sqrt{\frac{2}{N_r -1}}, \q\q
\La_{\th} = \pi \mathcal{P}^{(\th)}_0\sqrt{\frac{2}{N_\th -1}} .
\end{align}
Taken all together, these solutions for $\la^{(r)}(\psi)$ and
$\la^{(\th)}(\chi)$ end up accurately relating motion in $r$ and $\th$ with 
$\la$.  This approach models that found in Sec.~II of \cite{HoppETC15}.

We proceed next to find the motion in $t$ and $\vp$.  With \eqref{eqn:Phi} and
\eqref{eqn:T} reexpressed in terms of Mino time, the periodicity of
\eqref{eqn:PhiR}-\eqref{eqn:TTh}, and the ability to express those functions
in terms of $\la$, suggests a Mino-time Fourier decomposition of $T^{(r)}$,
$T^{(\th)}$, $\Psi^{(r)}$, and $\Psi^{(\th)}$ \cite{DrasHugh05}
\begin{align}
\label{eqn:PhiRseries}
\Psi^{(r)}(r_p(\la)) &= \sum_{n=-\infty}^{+\infty} \wp^{(r)}_n \,
e^{-in\Upsilon_r \la} ,
\\
\label{eqn:PhiThseries}
\Psi^{(\th)}(\th_p(\la)) &= \sum_{k=-\infty}^{+\infty} \wp^{(\th)}_k \,
e^{-ik\Upsilon_{\th}\la} ,
\\
\label{eqn:TRseries}
T^{(r)}(r_p(\la)) &= \sum_{n=-\infty}^{+\infty} \mathcal{T}^{(r)}_n \,
e^{-in\Upsilon_r \la} ,
\\
\label{eqn:TThseries}
T^{(\th)}(\th_p(\la)) &= \sum_{k=-\infty}^{+\infty} \mathcal{T}^{(\th)}_k \,
e^{-ik\Upsilon_{\th}\la} ,
\end{align}
where in keeping with the left-hand sides being real functions the
coefficients will satisfy crossing relations
(e.g., $\mathcal{T}^{(r)}_{-n} = \mathcal{T}^{(r)*}_n$).  As before, the
series might be truncated (here with some upper and lower bounds on $n$ and
$k$).  The Fourier coefficients are found from integrals over $\la$; for
example
\be
\label{eqn:fsTR}
\mathcal{T}^{(r)}_n = \frac{1}{\La_r} \int_0^{\La_r} T^{(r)} \,
e^{in\Upsilon_r \la} \, d\la ,
\ee
with similar integrals for $\mathcal{T}^{(\th)}_k$, $\wp^{(r)}_{n}$, and
$\wp^{(\th)}_{k}$.

If we introduced sufficiently fine, evenly spaced divisions of the respective
periods in $\la$, each of the Fourier coefficient integrals, like
\eqref{eqn:fsTR}, could be accurately replaced with a finite sum.
Unfortunately, the functions being integrated depend on $r_p$ or $\th_p$
(e.g., $T^{(r)}(r)$ above), which are known from the previous analysis as
functions sampled on evenly spaced grids in $\psi$ or $\chi$.  Rather than
resample them to an evenly spaced grid in $\la$, we instead convert the
integrals and integrate over $\psi$ or $\chi$.  For example
\begin{align}
\mathcal{T}^{(r)}_n &= \frac{1}{\La_r} \int_0^{2\pi} T^{(r)} \, P^{(r)} \,
e^{in\Upsilon_r \la^{(r)}(\psi)} \, d\psi ,
\\
\mathcal{T}^{(\th)}_k &= \frac{1}{\La_{\th}} \int_0^{2\pi} T^{(\th)} \,
P^{(\th)} \, e^{ik\Upsilon_{\th} \la^{(\th)}(\chi)} \, d\chi ,
\end{align}
with similar expressions for $\wp^{(r)}_{n}$ and $\wp^{(\th)}_{k}$.  Despite
the transformations, all of these integrands are still $C^{\infty}$ periodic
functions of (now) $\psi$ or $\chi$.  As was shown in \cite{HoppETC15}
(Sec.~III.B.3), smooth reparameterizations of this sort still allow
exponentially convergent approximations to be made by replacing an integral
with a finite sum on an evenly spaced grid in the new coordinate (either
$\psi$ or $\chi$)
\begin{align}
\label{eqn:psiPoints}
\psi_j &\equiv \frac{2 j \pi}{N_r},	  \q\q\q
j \in {0, 1, \ldots, N_r-1},
\\
\label{eqn:Tcoeffs}
\mathcal{T}^{(r)}_n &\simeq \frac{\Upsilon_r}{N_r}
\sum_{j=0}^{N_r-1}
T^{(r)}(\psi_j) \, P^{(r)}(\psi_j) \,
e^{in\Upsilon_r \la^{(r)}(\psi_j)} ,
\\
\label{eqn:chiPoints}
\chi_s &\equiv \frac{2 s \pi}{N_{\th}} , \q\q\q
s \in {0, 1, \ldots, N_{\th}-1},
\\
\label{eqn:Phicoeffs}
\mathcal{T}^{(\th)}_k &\simeq \frac{\Upsilon_{\th}}{N_{\th}}
\sum_{s=0}^{N_{\th}-1}
T^{(\th)}(\chi_s) \, P^{(\th)}(\chi_s) \,
e^{ik\Upsilon_{\th} \la^{(\th)}(\chi_s)} .
\end{align}
Similar expressions again hold for $\wp^{(r)}_{n}$ and $\wp^{(\th)}_{k}$.
Because \eqref{eqn:Tcoeffs} and \eqref{eqn:Phicoeffs} are not evaluated on an
evenly spaced, periodic grid in $\la$, they do not represent DFT sums (the
argument of the exponential is nonlinear in $\psi$ or $\chi$).  Accordingly,
the coefficients cannot be computed with the $\mathcal{O}(N \log N)$ FFT
algorithm, but instead are evaluated directly, which is an
$\mathcal{O}(N^2)$ process.

Once the Fourier coefficients are known, the average $\la$ accumulation
rates, $\Gamma $ and $\Upsilon_{\vp}$, are found from the leading coefficients
\begin{align}
\Gamma = \mathcal{T}^{(r)}_0+\mathcal{T}^{(\th)}_0, \q\q\q
\Upsilon_{\vp} = \wp^{(r)}_0+\wp^{(\th)}_0 .
\end{align}
The remaining parts that determine the advance of $t$ and $\vp$ in
\eqref{eqn:tpOfLa} and \eqref{eqn:phipOfLa}, the periodic functions $\D t_p$
and $\D \vp_p$, may be expressed as functions of $\la$ by integrating
\eqref{eqn:PhiRseries}-\eqref{eqn:TThseries} term-by-term
\begin{align}
\label{eqn:DtpOfLaR}
\D t^{(r)}(\la) &= 2\, \text{Re}\left[\sum_{n=1}^{N_r/2}
\frac{i\mathcal{T}^{(r)}_n}{n\Upsilon_r} \,
e^{-in\Upsilon_r \la} \right] ,
\\
\label{eqn:DtpOfLaTh}
\D t^{(\th)}(\la) &= 2\, \text{Re}\left[\sum_{k=1}^{N_{\th}/2}
\frac{i\mathcal{T}^{(\th)}_k}{k\Upsilon_{\th}} \,
e^{-ik\Upsilon_{\th} \la} \right] ,
\\
\label{eqn:DphipOfLaR}
\D \vp^{(r)}(\la) &= 2\, \text{Re}\left[\sum_{n=1}^{N_r/2}
\frac{i\wp^{(r)}_n}{n\Upsilon_r} \,
e^{-in\Upsilon_r \la} \right] ,
\\
\label{eqn:DphipOfLaTh}
\D \vp^{(\th)}(\la) &= 2\, \text{Re}\left[\sum_{k=1}^{N_{\th}/2}
\frac{i\wp^{(\th)}_k}{k\Upsilon_{\th}} \,
e^{-ik\Upsilon_{\th} \la} \right] .
\end{align}
Here $N_r$ and $N_\th$ are assumed to be even and the restricted range of
the sums reflects use of the crossing relations.

\subsection{Analytic mode functions from MST formalism}
\label{sec:MST}

The MST formalism \cite{ManoSuzuTaka96a,ManoSuzuTaka96b} ultimately provides
radial mode function solutions $\hat{X}^\pm_{\lhat mkn}$ to \eqref{eqn:genSN}
subject to the boundary conditions \eqref{eqn:Xp} and \eqref{eqn:Xm}.  The
formalism more traditionally yields the radial Teukolsky functions
$R_{\lhat m\o}$ (in our case spin weight equal zero), from which follow
$\hat{X}^\pm_{\lhat mkn}$.  A comprehensive review of the MST formalism is given
in \cite{SasaTago03}.  Our presentation here primarily focuses on efficient
calculation of one set of these solutions.  The calculation first starts by
determining the separation constant $\la_{\lhat mkn}$.  We make use of the
Black Hole Perturbation Toolkit's \cite{BHPTK18} \textit{Mathematica} package
\textsc{SpinWeightedSpheroidalHarmonics} to evaluate $\la_{\lhat mkn}$.  

The Teukolsky functions $R^\text{in}_{\lhat m\o}$ and
$R^\text{up}_{\lhat m\o}$ are the solutions to the radial Teukolsky equation
with boundary conditions
\begin{align}
\label{eqn:RinBC}
R^\text{in}_{\lhat m\o}(r\rightarrow r_+) &\simeq B^\text{trans}e^{-i\g r_*} ,
\\
\label{eqn:RupBC}
R^\text{up}_{\lhat m\o}(r\rightarrow\infty)
&\simeq C^\text{trans} r^{-1} e^{i\o r_*} ,
\end{align}
that correspond to the conditions \eqref{eqn:Xm} and \eqref{eqn:Xp},
respectively, on $\hat{X}^\pm_{\lhat mkn}$.  Here $B^\text{trans}$ and
$C^\text{trans}$ are asymptotic amplitudes. 
By introducing the renormalized angular momentum $\nu$ and rescaling the
radial coordinate in two convenient ways
\begin{equation}
  x = \frac{r_+-r}{2 M \kappa}, \qquad \qq z=\o(r-r_-) ,
\end{equation}
the functions $R^\text{in}_{\lhat m \o}$ and $R^\text{up}_{\lhat m \o}$ are 
expressed as series of hypergeometric functions,
\begin{align}
\label{eqn:mstRin}
  R^\text{in}_{\lhat m \o} &= e^{i \eps \kappa x}
  (-x)^{-i\eps_+}(1-x)^{i\eps_-} \\
  & \times \sum_{n=-\infty}^{+\infty} a^\nu_n \,
  F\big(L+1-i\tau,-L-i\tau; 1-2i\eps_+; x \big) \notag
\\
\label{eqn:mstRup}
  R^\text{up}_{\lhat m\o} &= e^{iz}z^{\nu+i\eps_+}
  (z-\eps\kappa)^{-i\eps_+} \\
  & \q \times \sum_{n=-\infty}^{+\infty} b^\nu_n \, (2iz)^n \,
  \Psi\l L+1-i\eps,2L+2;-2iz\r, \notag
\end{align}
where here we have adopted $s=0$ (spin weight of the scalar case), which
we maintain henceforth in this paper.  Other parameters are
\begin{align}
L&= n+\nu, \qquad \eps=2 M \o, \qquad \kappa=\sqrt{1-\frac{a^2}{M^2}}, \notag
\\
&\tau = \frac{1}{\kappa}\l\eps - \frac{m a}{M}\r, \qq
\eps_\pm=\frac{1}{2}(\eps\pm\tau) .
\end{align}
In the expressions above, $F(c_1,c_2;c_3;x)$ is the Gauss hypergeometric
function $\mbox{}_2 F_1(c_1,c_2;c_3;x)$ and $\Psi(c_1,c_2;z)$ is the
irregular confluent hypergeometric function.

The series coefficients $a^\nu_n$ are the minimal solution to a three-term
recurrence relation that allows the series to converge once 
$\nu$ is determined.  The second set of coefficients $b^\nu_n$ are completely
determined by $a^\nu_n$ via
\begin{equation}
b^\nu_n = e^{-i\pi(\nu+1-i\eps)}2^{\nu}
\frac{(\nu+1-i\eps)_n}{(\nu+1+i\eps)_n} a^\nu_n ,
\end{equation}
making the ``up'' series convergent also.  Here 
$(\mu)_n \coloneqq \G(\mu+n)/\G(\mu)$
is the Pochhammer symbol.  For $n=-1,0,1$, we calculate $F(c_1,c_2;c_3;x)$ and
$\Psi(c_1,c_2;z)$ using \textit{Mathematica}'s built-in functions
\texttt{Hypergeometric2F1} and \texttt{HypergeometricU}, respectively.  For
$|n|>1$, we construct both types of hypergeometric functions using their
respective three-term recursion relations (provided in \cite{SasaTago03}).

The eigenvalue $\nu$ is often determined by solving for the root of a
complex equation with coefficients that are built from continued fractions
\cite{SasaTago03}.  An alternative method, employed in this paper, relates
$\nu$ to the eigenvalue of the monodromy matrix defined for the irregular
singular point of the Teukolsky equation at $r\rightarrow\infty$
\cite{CastETC13a,CastETC13b}.  Then $\nu$ is determined numerically by
calculating Stokes multipliers \cite{Rodr,CastETC13b}.

An accuracy goal in determining the radial functions is met in part by
terminating the hypergeometric series at a sufficiently large value of
$|n| = n_\text{max}$ (where $n_\text{max}$ is not necessarily the same for
both series).  The MST technique provides precise, semianalytic solutions,
but it can be computationally expensive, especially when programmed in
\textit{Mathematica}.  As the frequency increases, the hypergeometric series
expansions must range over an increasing number of terms to meet a
pre-defined accuracy goal.  Computational costs are exacerbated by roundoff
errors from near cancellations in the sums.  Roundoff errors are
circumvented by making internal \textit{Mathematica} calculations at working
precisions significantly higher than desired accuracy in final results.

We found empirically that, for the radial positions considered in this work, 
the series of confluent hypergeometric functions
$\Psi(c_1,c_2;z)$ converges more rapidly than the series of Gauss
hypergeometric functions $F(c_1,c_2;c_3;x)$ (used in the ``in'' solution).
Further study showed that computational costs can be mitigated on the
horizon side in calculating $R^\text{in}$ by using an alternative expression
given in the MST literature (see Eq.~(166) in \cite{SasaTago03})
\begin{equation}
\label{eqn:RinAndRnuC}
R^\text{in} = K_\nu R^\nu_\text{C} + K_{-\nu-1} R^{-\nu-1}_\text{C},
\end{equation}
where $R^\nu_\text{C}$ is expressed as a series of regular confluent
hypergeometric functions $M(c_1,c_2;z)$,
\begin{multline}
\label{eqn:RnuC}
R^\nu_\text{C} = e^{-iz} z^{\nu+i\eps_+}\l z-\eps\kappa
\r^{-i\eps_+}
\\
\times \sum_{n=-\infty}^{+\infty} f^\nu_n \, (-2iz)^n \, M(L+1+i\eps,2L+2;2iz) .
\end{multline}
Here $f^\nu_n$ is a new set of series coefficients (given below) and
$K_\nu$ is a (phase) factor that involves summing over the prior series
coefficients $a^\nu_n$ and $b^\nu_n$.  The exact form of $K_\nu$ in our case
is given by
\begin{multline}
K_\nu = e^{i\eps\kappa}(\eps\kappa)^{-\nu} \, \G(1-2i\eps_+) \, \G(2\nu+1)
\\
\times \l \sum_{n=0}^{+\infty} \frac{(-1)^n}{n!} g^\nu_n \r \l \sum_{n=-\infty}^0 \frac{(-1)^n}{(-n)!} h^\nu_n\r^{-1} .
\end{multline}
The new series coefficients $f^\nu_n$, $g^\nu_n$, and $h^\nu_n$ can be
expressed in terms of the prior coefficients $a^\nu_n$ and $b^\nu_n$ by
\begin{align}
f^\nu_n &= e^{i\pi(\nu+1-i\eps)} \frac{\G (L+1+i\eps)}{\G(2L+2)} b^\nu_n,
\\
g^\nu_n &= (2\nu+1)_n \frac{(\nu+1+i\tau)_n}{\G(L+1-i\tau)}
\frac{(\nu+1+i\eps)_n} {\G(L+1-i\eps)} a^\nu_n
\\
h^\nu_n &= e^{i\pi(\nu+1-i\eps)} \frac{\G(L+1+i\eps-n)}{\G(2L+2-n)}
b^\nu_n.
\end{align}

The review by Sasaki and Tagoshi \cite{SasaTago03} discusses
\eqref{eqn:RinAndRnuC} as a complement of \eqref{eqn:mstRin} that
provides convergent coverage of the entire domain $[r_{+},+\infty]$, but
does not mention its computational efficiency.  Rapid convergence was our focus
in comparing these expressions and settling on use of \eqref{eqn:RinAndRnuC}.
While writing this paper, we sought other MST users' experiences with the
potential practical virtues of \eqref{eqn:RinAndRnuC} and \eqref{eqn:RnuC}.
Casals \cite{BussCasa18} and Wardell (private communication) were aware of
the benefits of \eqref{eqn:RinAndRnuC} and make use of it in their work,
though have not previously discussed this particular issue in detail.  Use of
both \eqref{eqn:RinAndRnuC} and \eqref{eqn:mstRin} are described by Throwe
\cite{Thro10}, with his observation that both formulae have their own regions 
in which they are numerically more suitable.  Elsewhere \cite{BHPTK18} 
Eq.~\eqref{eqn:mstRin} is exclusively used.

A side benefit in our approach is that the series of regular confluent
hypergeometric functions $M(c_1,c_2;z)$ given in \eqref{eqn:RnuC} converges
with similar rapidness as the series of irregular confluent hypergeometric
functions $\Psi(c_1,c_2;z)$ given in \eqref{eqn:mstRup}.  Thus the same value
of $n_\text{max}$ can be used to truncate both series.

While use of $R_C$ has benefits, it is not straightforward to construct
the underlying functions $M(c_1,c_2;z)$ numerically.  The functions
$M(c_1,c_2;z)$ satisfy a three-term recurrence relation \cite{SasaTago03} but
evaluating the functions by stepping through the recurrence formula is
numerically unstable in the increasing-$n$ direction.  There are several ways
to circumvent this problem: increase the code's internal precision, calculate
$M(c_1,c_2;z)$ directly using \textit{Mathematica}'s built-in function
\texttt{Hypergeometric1F1}, or translate the three-term recurrence relation
into a continued fraction, which does not suffer from the same cancellation
errors in the increasing-$n$ direction.  Alternatively, since the recurrence
relation does not suffer the same instability when moving down in $n$, one
can begin the summation of \eqref{eqn:RnuC} at $n=n_\text{max}$ and evaluate
the terms as $n$ decreases down to $n=-n_\text{max}$.  The value of
$n_\text{max}$ is conveniently determined by evaluating $R^\text{up}$ first.
A mixture of these strategies is employed to maximize computational efficiency.
Ultimately the improved convergence of \eqref{eqn:RinAndRnuC} and
\eqref{eqn:RnuC}, compared to \eqref{eqn:mstRin}, offsets the computational
cost of summing two series instead of one.

Using these expressions for $R^\text{in}$ and $R^\text{up}$, we can construct
the unit-normalized functions $\hat{X}^\pm$ by comparing \eqref{eqn:RinBC} and
\eqref{eqn:RupBC} with \eqref{eqn:SNtrans}, \eqref{eqn:Xp}, and \eqref{eqn:Xm}
\be
\hat{X}^- = \frac{\varpi}{\varpi_+}\l\frac{R^\text{in}}{B^\text{trans}}\r ,
\quad\quad
\hat{X}^+ = \varpi\l\frac{R^\text{up}}{C^\text{trans}}\r ,
\ee
where $\varpi_+ = (r_+^2+a^2)^{1/2}$.  The asymptotic amplitudes can be 
found by expanding both solutions near the horizon and at large $r$
\begin{align}
B^\text{trans} & = e^{i\kappa\eps_+
\l 1+\frac{2\ln\kappa}{1+\kappa}\r } \sum_{n=-\infty}^{+\infty} a^\nu_n ,
\\
C^\text{trans} & = \o^{-1}
e^{i(\eps\ln\eps-\frac{1-\kappa}{2}\eps)}A^\nu_- ,
\end{align}
with
\begin{equation}
A^\nu_- = 2^{-(\nu+1-i\eps)} e^{i\pi(\nu+1-i\eps)/2} \sum^{+\infty}_{n=-\infty} (-1)^n \, b^\nu_n.
\end{equation}

\subsection{Optimized source integration}
\label{sec:kerrSSI}

\begin{figure}[tbp]
\includegraphics[width=0.48\textwidth]{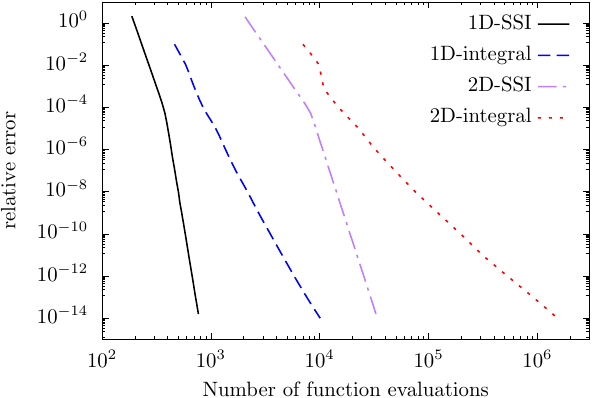}
\caption{Computational efficiency in calculating normalization coefficients.
An assessment of computational efficiency is made by measuring the number of
integrand evaluations needed to calculate $C^+_{2222}$ and $C^-_{2222}$ for
orbital parameters $(p,e,\iota,a/M)=(15,0.5,\pi/3,0.5)$.  The lowest
efficiency and slowest convergence rate is that of the 2D-integral approach
(red dotted curve).  The effect of switching to products of 1D integrals is seen in
the 1D-integral method (blue dashed curve).  The effect of switching from
adaptive-step integration to SSI is seen in the 2D-SSI (purple dot-dashed) and 1D-SSI
(black solid) scalings.  The adaptive step-size integrations (both 2D-integral and
1D-integral) converge algebraically at 8th order.}
\label{fig:SSI}
\end{figure}

We consider next the optimized calculation of the normalization coefficients
$C^\pm_{\lhat mkn}$ defined in \eqref{eqn:Z}.  That reduction begins with a
review of the derivation of the FD source function $Z_{\lhat mkn}(r)$,
exploiting the orthogonality of the harmonics in $t$ and $\vp$, and the
spheroidal Legendre functions found in \eqref{eqn:sourceDecomp}.  Integrating
the product of \eqref{eqn:sourceDecomp} and $e^{-im\vp}$ over azimuth angle
and using the delta function in $\vp$, we find
\begin{align}
&\sum_{\lhat kn} Z_{\lhat mkn}(r)\, S_{\lhat mkn}(\th)\, e^{-i\o_{mkn} t}
\\
&\qquad \qquad \; = -\frac{2 q \Sig \D\,\d(r-r_p)\,
\d(\cos\th-\cos\th_p)}{ \varpi^3\left( T^{(r)}+T^{(\th)} \right)} \,
e^{-im\vp_p} . \notag
\end{align}
We next remove the linear phase factor $e^{-im\O_\vp t}$, which makes the
remaining expression
\begin{widetext}
\be
\label{eqn:lknSum2}
\sum_{\lhat kn} Z_{\lhat mkn}(r)\, S_{\lhat mkn}(\th)\,
e^{-i(k\O_\th+n\O_r) t}
= -2 q \frac{e^{-im\left(\D\vp^{(r)}+\D\vp^{(\th)}
-\O_\vp(\D t^{(r)}+\D t^{(\th)}) \right)}}
{\varpi_p^3\left( T^{(r)}+T^{(\th)} \right)}
\Sig_p\, \D_p \d(r-r_p)\, \d(\cos\th-\cos\th_p) ,
\ee
biperiodic with fundamental frequencies $\O_\th$ and $\O_r$, since
$\vp_p-\O_\vp t = \D\vp^{(r)}+\D\vp^{(\th)}-\O_\vp(\D t^{(r)}+\D t^{(\th)})$
up to an irrelevant constant.

We next reduce \eqref{eqn:lknSum2} to a single sum over $\hat{l}$ by using
orthogonality of the factor $e^{-i(k\O_\th+n\O_r) t}$.  To do so, we convert
to Mino time Fourier series, with $e^{-i(k\Upsilon_\th+n\Upsilon_r) \lambda}$,
using results in \cite{DrasHugh05}
\be
\label{eqn:lSum}
\sum_{\lhat} Z_{\lhat mkn}(r)\, S_{\lhat mkn}(\th)
= \frac{1}{\La_\th \La_r}\int\limits_0^{\La_\th}  d\la^{(\th)}
\int\limits_0^{\La_r} d\la^{(r)} \,
e^{i(k\Upsilon_\th \la^{(\th)}+n \Upsilon_r \la^{(r)})}
B_{mkn}(r_p,\th_p) \, \d(r-r_p)\, \d(\cos\th-\cos\th_p) ,
\ee
where the function $B_{mkn}(r_p,\th_p)$ is
\begin{equation}
B_{mkn}(r_p,\th_p) \equiv -\frac{4\pi q \Sig_p \D_p}{ \G\varpi_p^3} \,
e^{i\o_{mkn}(\D t^{(r)}+\D t^{(\th)})} \, e^{-im(\D\vp^{(r)}+\D\vp^{(\th)})} ,
\end{equation}
which can be thought of as a function of $\la^{(r)}$ and $\la^{(\th)}$.  The
final step in deriving $Z_{\lhat mkn}(r)$ is multiplying the above expression
by $S_{\lhat mkn}(\th)$ and integrating over $\th$
\be
\label{eqn:Zexpr}
Z_{\lhat mkn}(r)
= \frac{1}{\La_\th \La_r}\int\limits_0^{\La_\th}  d\la^{(\th)}
\int\limits_0^{\La_r} d\la^{(r)} \,
e^{i(k\Upsilon_\th \la^{(\th)}+n \Upsilon_r \la^{(r)})} B_{mkn}(r_p,\th_p)
\, S_{lmkn}(\th_p) \, \d(r-r_p) .
\ee

With the FD source function in hand, we may calculate the normalization
constants $C^\pm_{\lhat mkn}$ by substituting \eqref{eqn:Zexpr} into
\eqref{eqn:Z}
\begin{equation}
C^\pm_{\lhat mkn} = \frac{1}{W_{\lhat mkn}}
\int\limits_{r_\text{min}}^{r_\text{max}} dr
\frac{\varpi^2 \hat{X}^\mp_{\lhat mkn}(r)}{\D }
\frac{1}{\La_\th \La_r}
\int\limits_0^{\La_\th}  d\la^{(\th)}
\int\limits_0^{\La_r} d\la^{(r)} \,
e^{i(k\Upsilon_\th \la^{(\th)}+n \Upsilon_r \la^{(r)})}
B_{mkn}(r_p,\th_p) \, S_{lmkn}(\th_p) \, \d(r-r_p) .
\end{equation}
The order of integration is exchanged, allowing the $r$ integral to be
evaluated first
\be
\label{eqn:2Dint}
C^\pm_{\lhat mkn} =
\frac{1}{\La_\th \La_r}
\int\limits_0^{\La_\th}  d\la^{(\th)}
\int\limits_0^{\La_r} d\la^{(r)} \,
e^{i(k\Upsilon_\th \la^{(\th)}+n \Upsilon_r \la^{(r)})} \,
D^\pm_{\hat{l}mkn}(r_p,\th_p) ,
\ee
where $D^\pm_{\hat{l}mkn}(r_p,\th_p)$, implicitly a function of $\la^{(r)}$
and $\la^{(\th)}$, is given by
\be
\label{eqn:Dlmkn}
D^\pm_{\hat{l}mkn}(r_p,\th_p) =
-\frac{4 \pi q \Sig_p \hat{X}^\mp_{\lhat mkn}(r_p) \,
S_{lmkn}(\th_p)}{\G\,W_{\lhat mkn}\, \varpi } \,
e^{i\o_{mkn}(\D t^{(r)}+\D t^{(\th)})}e^{-im(\D\vp^{(r)}+\D\vp^{(\th)})} .
\ee

The double integral in \eqref{eqn:2Dint} may be computed directly using
adaptive-step-size integration \cite{DrasHugh06}.  We refer to this method
henceforth as the ``2D-integral'' approach, which can be shown to deliver
numerical results that converge algebraically (i.e., as a power law).  Given
the number of modes in
the Kerr generic-orbit problem, this is a computationally expensive method
that compelled us to search for more efficient alternatives in evaluating
\eqref{eqn:2Dint}.

A first alternative is to exploit the integrand's smoothness and
bi-periodicity to make a discrete, evenly spaced sampling in two dimensions
that is analogous to the approach we took with the orbit equations.  Just as
in that case, where an equally spaced sum over samples of a smooth periodic
integrand converged exponentially, we find ``spectral'' convergence in the
two-dimensional integral as well.  Using the discrete sampling locations
of \eqref{eqn:psiPoints} and \eqref{eqn:chiPoints}, we calculate
\be
C^{\pm}_{\hat{l}mkn} \simeq \frac{\Upsilon_r\Upsilon_\th}{N_r N_\th}
\sum_{j=0}^{N_r-1} \sum_{s=0}^{N_\th-1}
e^{in\Upsilon_r\la^{(r)}(\psi_j)} \,
e^{ik\Upsilon_{\th}\la^{{(\th)}}(\chi_s)}
P^{(r)}(\psi_j)\,P^{(\th)}(\chi_s) \,
D^\pm_{\hat{l}mkn}(r_{p,j},\th_{p,s}) ,
\label{eqn:2Dssi}
\ee
\end{widetext}
where we have changed the integration variables from $\la^{(r)}$ and
$\la^{(\th)}$ to $\psi$ and $\chi$, adopted $r_{p,j} \equiv r_p(\psi_j)$
and $\th_{p,s} \equiv \th_p (\chi_s)$, and let the arguments of $D$ reflect
the discrete sampling.  The integration approach in
\eqref{eqn:2Dssi} is referred to here as the ``2D-SSI'' method (i.e.,
the two-dimensional generalization of the SSI technique \cite{HoppETC15}).
Figure \ref{fig:SSI} demonstrates the increased efficiency of the 2D-SSI method
compared to the 2D-integral scheme.  The 2D-SSI method has been independently
adopted by van de Meent \cite{Vand17} in his GSF FD calculations on inclined
eccentric orbits in Kerr spacetime.  We also understand that the code used
in \cite{DrasHugh06} has been upgraded to use the 2D-SSI method (Hughes,
private communication).

The explicit dependence on $r_p$ and $\th_p$ found in \eqref{eqn:Dlmkn}
allows for further optimization.  Because $D^\pm_{\hat{l}mkn}(r_p,\th_p)$
can be written in the following form
\be
\label{eqn:DJK}
D^\pm_{\hat{l}mkn} = \left(r_p^2 +a^2 \cos^2{\th_p} \right)
J_{\hat{l}mkn}(\th_p) \, K^\pm_{\hat{l}mkn}(r_p) ,
\ee
\be
J_{\hat{l}mkn}(\th_p) \equiv \frac{4 \pi q}{\G}
S_{lmkn}(\th_p) e^{i\o_{mkn}\D t^{(\th)}}e^{-im\D\vp^{(\th)}} ,
\ee
\be
K^\pm_{\hat{l}mkn}(r_p) \equiv
-\frac{\hat{X}^\mp_{\lhat mkn}(r_p)}{W_{\lhat mkn}\, \varpi} \,
e^{i\o_{mkn}\D t^{(r)}}e^{-im\D\vp^{(r)}} ,
\ee
the double integral in \eqref{eqn:2Dint} can be calculated from products of
one-dimensional integrals
\begin{align}
\label{eqn:C1D}
&C^\pm_{\lhat mkn} = I^{(1)\pm}_{\lhat mkn} \, I^{(2)}_{\lhat mkn}
+ I^{(3)\pm}_{\lhat mkn} \, I^{(4)}_{\lhat mkn},
\\
\label{eqn:I1pm}
&I^{(1)\pm}_{\lhat mkn} \equiv
\frac{1}{\La_r} \int_0^{\La_r}  d\lambda^{(r)} \,
e^{i n \Upsilon_r \lambda^{(r)}} r_p^2 K^\pm_{\lhat mkn}(r_p),
\\
\label{eqn:I2}
&I^{(2)}_{\lhat mkn} \equiv
\frac{1}{\La_\th} \int_0^{\La_\th}  d\lambda^{(\th)} \,
e^{i k \Upsilon_\th \lambda^{(\th)}} J_{\lhat mkn}(\th_p),
\\
\label{eqn:I3pm}
&I^{(3)\pm}_{\lhat mkn} \equiv
\frac{1}{\La_r} \int_0^{\La_r}  d\lambda^{(r)} \,
e^{i n \Upsilon_r \lambda^{(r)}} K^\pm_{\lhat mkn}(r_p),
\\
\label{eqn:I4}
&I^{(4)}_{\lhat mkn} \equiv
\frac{a^2}{\La_\th} \int_0^{\La_\th}  d\lambda^{(\th)}
e^{i k \Upsilon_\th \lambda^{(\th)}} \cos^2\th_p J_{\lhat mkn}(\th_p).
\end{align}
If we use these equations and just compute the integrals
\eqref{eqn:I1pm}-\eqref{eqn:I4} with a straightforward adaptive integrator,
we get an algebraically convergent method that we refer to as the
``1D-integral'' approach.  Despite its algebraic convergence, it is much
faster at any required level of accuracy than the 2D-integral approach, by
as much as two orders of magnitude at conventional double precision (as seen
in Fig.~\ref{fig:SSI}).  At that accuracy level it is also faster than 2D-SSI,
though the faster convergence rate of 2D-SSI would ultimately win at higher
accuracies.

Finally, the 1D integrals are just as amenable to the SSI method as the
double integral and it is possible to make an exponentially convergent
discrete representation for \eqref{eqn:I1pm}-\eqref{eqn:I4}
\begin{align}
&\psi_j \equiv \frac{2 j \pi}{N_{1,3}},\q\q\q j \in {0, 1, \ldots, N_{1,3}-1},
\notag
\\
&\chi_s \equiv \frac{2 s \pi}{N_{2,4}} ,\q\q\q s \in {0, 1, \ldots, N_{2,4}-1},
\notag
\\
&I^{(1)\pm}_{\lhat mkn} \simeq \frac{\Upsilon_r}{N_1}
\sum_{j=0}^{N_1-1} e^{i n\Upsilon_r \la^{(r)}(\psi_j)} P^{(r)}(\psi_j)
\label{eqn:I1DFT}
\\
&\qquad\qquad\qquad\qquad\;\;\; \times r_{p,j}^2 \,
K^\pm_{\lhat mkn}\big(r_{p,j} \big),
\notag
\\
&I^{(2)}_{\lhat mkn} \simeq \frac{\Upsilon_\th}{N_2}
\sum_{s=0}^{N_2-1}e^{i k\Upsilon_{\th} \la^{(\th)}(\chi_s)} P^{(\th)}(\chi_s)
\, J_{\lhat mkn}\big(\th_{p,s}\big),
\\
&I^{(3)\pm}_{\lhat mkn} \simeq \frac{\Upsilon_r}{N_3}
\sum_{j=0}^{N_3-1} e^{i n\Upsilon_r \la^{(r)}(\psi_j)} P^{(r)}(\psi_j) \,
K^\pm_{\lhat mkn}\big(r_{p,j}\big),
\\
&I^{(4)}_{\lhat mkn} \simeq \frac{\Upsilon_\th}{N_4}
\sum_{s=0}^{N_4-1}e^{i k\Upsilon_{\th} \la^{(\th)}(\chi_s)} P^{(\th)}(\chi_s)
\label{eqn:I4DFT}
\\
&\qquad\qquad\qquad\qquad\;\;\; \times a^2\cos^2\th_{p,s} \,
J_{\lhat mkn}\big(\th_{p,s}\big) .
\notag
\end{align}
When \eqref{eqn:I1DFT}-\eqref{eqn:I4DFT} are used to evaluate \eqref{eqn:C1D},
we refer to it as the ``1D-SSI'' method.  Figure \ref{fig:SSI} shows that the
1D-SSI method is the most efficient and most rapidly convergent technique.
Switching to 2D-SSI from 2D adaptive-step integration is nearly two orders
of magnitude faster at double precision accuracies.  Switching from 2D-SSI to
1D-SSI yields another factor of 30.

The 1D-SSI method is possible because the two-dimensional source integrations
decompose as shown in \eqref{eqn:C1D} into products of 1D integrals.
Unfortunately a similar decomposition does not occur in any obvious way for
gravitational perturbations in Kerr spacetime due to leading factors of
$1/\Sig$.  For small spins or large radial separations, the $1/\Sig$ factor
might be expanded using a binomial series with a modest amount of terms,
providing an \emph{approximately} separable source.  It is also conceivable
that a transformation might exist that would bring the source into a separable
form.  The benefits of the 1D-SSI method seen in the scalar case are
compelling enough to justify a more thorough investigation of the
gravitational Teukolsky source integration problem.

\section{Generic Orbit SSF Regularization}
\label{sec:regularize}

\subsection{Mode-sum regularization review}
\label{sec:modeSumReg}

Section \ref{sec:teuk} provides a roadmap for calculating the retarded field,
$\Phi^\text{ret}$, including its decomposition in a spherical harmonic basis,
and Sec.~\ref{sec:ssf} discusses using the gradient of that field and the
singular field (with the vector components also expanded in the same basis)
to yield the mode-sum regularized self-force
\begin{align}
\label{eqn:ssfModeSum2}
F_{\a} = \sum_{l=0}^{+\infty}
\left( F^{\text{ret},l}_{\a\pm} - F^{\text{S},l}_{\a\pm} \right) .
\end{align}
This equation differs from
\eqref{eqn:ssfModeSum} in making clear that individual $l$-mode self-force
components may differ in value in the limit as $r\rightarrow r_p$ depending
upon the direction of approach in $r$.  This $\pm$ notation aligns with that
used in the EHS discussion (i.e., Eq.~\eqref{eqn:EHS2}) of mode functions.  
Using the spherical harmonic decomposition \eqref{eqn:phiSpherical} of the 
retarded field, the $l$-modes of three of the force components are
\begin{align}
&F^{\text{ret},l}_{t\pm} = q \lim_{x \rightarrow x_p}
\sum_{m=-l}^{l} \partial_t\phi^\pm_{lm}(t,r)\, Y_{lm}(\th,\vp) ,
\\
&F^{\text{ret},l}_{r\pm} = q \lim_{x \rightarrow x_p}
\sum_{m=-l}^{l} \partial_r\phi^\pm_{lm}(t,r)\, Y_{lm}(\th,\vp) ,
\\
&F^{\text{ret},l}_{\vp\pm} = q \lim_{x \rightarrow x_p}
\sum_{m=-l}^{l} im\,\phi^\pm_{lm}(t,r) \, Y_{lm}(\th,\vp) .
\end{align}
The $\th$ component\footnote{In the GSF case it is
sufficient to regularize just three of the four force components because the
final component is fixed by $u^\a F_\a = 0$.  In the SSF case the force has a
tangential component along $u^\a$, leading to variation in mass
\cite{QuinWald97,Quin00,BurkHartPois02,DrasFlanHugh05,PoisPounVega11} and
requiring calculation and regularization of $F_{\th}$.} of the self-force
is broken down into $l$-modes, $F^{\text{ret},l}_{\th\pm}$, only after the
derivative $\pa_\th Y_{lm}$ is reprojected onto the $Y_{lm}$ basis.

To effect this change, we use the clever window function $f(\th)$ devised by
Warburton \cite{Warb15} (his Eq.~50).  When multiplied with the field, $f(\th)$
affects neither the value of the field as $\th \rightarrow \th_p$ nor its
first derivative, yet produces a finite coupling between
$f(\th)\,\partial_\th Y_{lm}$ and (up to) four spherical harmonics $Y_{lm}$.
This allows the $l$-modes of the $\th$-component to be reexpressed as
\begin{align}
\label{eqn:fthYlm}
&F^{\text{ret},l}_{\th\pm} = q \lim_{x \rightarrow x_p}
\sum_{m=-l}^{l} \psi^\pm_{lm}(t,r) \, Y_{lm}(\th,\vp) ,
\end{align}
where the $\psi^\pm_{lm}(t,r)$ are defined in terms of $\phi^\pm_{lm}(t,r)$
using the following condensed notation
\begin{multline}
\label{eqn:fthExpansion}
\psi^\pm_{lm} = \b_{l+3,m}^{(-3)}\,\phi^\pm_{l+3,m}
+ \b_{l+1,m}^{(-1)}\,\phi^\pm_{l+1,m}
\\
+ \b_{l-1,m}^{(+1)}\,\phi^\pm_{l-1,m}
+ \b_{l-3,m}^{(+3)}\,\phi^\pm_{l-3,m} .
\end{multline}
The coefficients $\b_{lm}^{(\pm i)}$, and a more detailed discussion of
deriving Eqs.~\eqref{eqn:fthYlm} and \eqref{eqn:fthExpansion}, are provided
in Appendix \ref{app:thetaReg}.  Our expressions \eqref{eqn:fthYlm} and
\eqref{eqn:fthExpansion} are similar to ones found in \cite{Warb15} with the
exception of minor corrections.

To calculate $F_{\a}$ from Eq.~\eqref{eqn:ssfDW} we require an expansion of
$F^{\text{S},l}_{\a\pm}$ in terms of regularization parameters
\cite{BaraOri00,BaraOri03a,DetwMessWhit03}
\begin{multline}
\label{eqn:singularForce}
F^{\text{S},l}_{\a\pm} = A_{\a\pm}L + B_\a
+ \sum_{n=1}^{+\infty} \frac{D_{\a,2n} }
{ \prod_{k=1}^n (2L-2k)(2L+2k) },
\end{multline}
where $L\equiv l+1/2$ and the parameters $A_{\a\pm}$, $B_\a$, and $D_{\a,2n}$
are all independent of $l$.  For each $n$, the higher-order regularization
terms (with coefficients $D_{\a,2n}$) have the property that the $l$-dependent
terms sum to zero \cite{DetwMessWhit03}:
\be
\sum_{l=0}^{+\infty} \left[ \prod_{k=1}^n (2L-2k)(2L+2k) \right]^{-1} = 0 .
\ee
As a consequence only the first two regularization parameters, $A_{\a\pm}$
and $B_{\a}$, are necessary to assure a convergent result and the regularized
self-force can be calculated from just
\be
\label{eqn:ssfRegABC}
F_{\a} = \sum_{l=0}^{+\infty}
\Big( F^{\text{ret},l}_{\a\pm} - A_{\a\pm}L-B_\a \Big)
\equiv \sum_{l=0}^{+\infty}  F^{\text{alg},l}_{\a\pm} ,
\ee
where we have defined $F^{\text{alg},l}_{\a\pm}$ for later convenience.
While the sum in \eqref{eqn:ssfRegABC} gives a finite result, the higher-order
terms drop off at a rate of $l^{-2}$.  When the sum is approximated by being
truncated at $l=l_\text{max}$, there is a residual error that scales as
$l_\text{max}^{-1}$.  Due to computational costs, it is typically
beneficial to truncate the SSF calculation at $l_\text{max}\sim 20$, which
means that relying only upon the regularization parameters $A_{\a\pm}$ and
$B_\a$ will determine $F_\a$ to just one or two digits of accuracy.

Including the higher-order parameters $D_{\a,2n}$ can improve the rate of
convergence of the partial sums of Eq.~\eqref{eqn:ssfModeSum2}, which are
now written as
\begin{equation}
\label{eqn:ssfRegABCD}
F_\a = \sum_{l=0}^{l_\text{max}}
\Bigg( F^{\text{alg},l}_{\a\pm} -\sum_{n=1}^{n_\text{max}}
\frac{D_{\a,2n} }{ \prod_{k=1}^n (2L-2k)(2L+2k) } \Bigg) .
\end{equation}
Here there is a two-fold truncation, with $l_\text{max}$ determining the
number of modes we calculate in the retarded field, $\Phi$, and
$n_\text{max}$ setting the limit in the number of available higher-order
regularization parameters.  Eq.~\eqref{eqn:ssfRegABCD} converges at a rate of
$l^{-2(n_\text{max}+1)}$ and therefore the SSF has an error that scales as
$l_\text{max}^{-2n_\text{max}-1}$.  Unfortunately, only $A_{\a\pm}$ and $B_\a$
are known analytically for generic orbits in Kerr \cite{BaraOri03a} (although,
terms up to $n_\text{max}=2$ are known for equatorial orbits in Kerr
\cite{HeffOtteWard14}).

We overcome the lack of analytically known higher-order regularization
parameters by fitting \cite{DetwMessWhit03} the high-$l$ contributions to
the SSF to the assumed form in \eqref{eqn:singularForce}, similar to the
means discussed in Sec.~IVC of Warburton and Barack \cite{WarbBara10}.
At high $l$, the self-force contributions are primarily determined by the
missing regularization parameters
\begin{equation}
\label{eqn:fitHigherRPs}
F^{\text{alg},l}_{\a\pm} \simeq \sum_{n=1}^{N}
\frac{D_{\a,2n} }{ \prod_{k=1}^n (2L-2k)(2L+2k) } .
\end{equation}
The number of regularization parameters $N$ that can be determined is limited
by the precision of $F^{\text{alg},l}_{\a\pm}$ and $l_\text{max}$.
We take the last $\bar{n}$ self-force $l$-mode contributions,
$F^{\text{alg},l}_{\a\pm}$, and fit these values to $N$ regularization
parameters by applying a least squares algorithm to
Eq.~\eqref{eqn:fitHigherRPs}.  The value of $\bar{n}$ is varied and a weighted
average is taken as described in \cite{WarbBara10}.  We also vary $N$ and use
the standard deviation of the results to estimate the error produced by this
fitting scheme.  However, we do not use Eq.~(47) in \cite{WarbBara10}, but
instead reapply the fitted regularization parameters using
Eq.~\eqref{eqn:ssfRegABCD} to improve the convergence of our SSF results.
The estimated errors are also propagated to determine the accuracy of the
SSF results.  Errors due to fitting typically dominate over the error from
terminating the $l$-mode summation.  The validity of these fits and their
errors is further discussed in Sec.~\ref{sec:schwOrbits}, where we compare 
fitted conservative self-force data (for an inclined Schwarzschild orbit) to 
conservative self-force data that has been regularized with known higher-order
parameters (for an equatorial Schwarzschild orbit).

\subsection{Conservative and dissipative self-force for generic orbits}
\label{sec:consDispSSF}

As mentioned in Sec.~\ref{sec:ssf}, the self-force can be decomposed into
dissipative and conservative components, $F^\text{diss}_\a$ and 
$F^\text{cons}_\a$,
which have different physical effects on the orbital evolution
\cite{Mino03,DiazETC04,Bara09}.  Just as we defined the retarded force
$F^\text{ret}_\a$, we similarly define the advanced force $F^\text{adv}_\a$ 
from the advanced scalar field solution, along with its $l$-mode contributions
$F^{\text{adv},l}_\a$.  Using the mode-sum scheme, the dissipative and
conservative components to the self-force are constructed from symmetric and
antisymmetric combinations of $F^{\text{ret/adv},l}_\a$
\begin{align}
F^{\text{diss}}_\a &= \sum^{+\infty}_{l=0} \frac{1}{2}
\l F^{\text{ret},l}_\a - F^{\text{adv},l}_\a \r ,
\\
F^{\text{cons}}_\a &= \sum^{+\infty}_{l=0} \left\{ \frac{1}{2}
\l F^{\text{ret},l}_\a + F^{\text{adv},l}_\a \r - F^{\text{S},l}_\a \right\} .
\end{align}
This decomposition is also beneficial for testing the numerical convergence of
the self-force results: the dissipative component does not need to be
regularized and will converge exponentially, while the conservative component
requires regularization and will converge algebraically as discussed in
Sec.~\ref{sec:modeSumReg}.

As is well known \cite{Mino03,HindFlan08}, the advanced and retarded forces
may both be obtained from the retarded solution, being related at 
\textit{reflection point} pairs in the orbital motion--points where the 
particle passes through the same radial and polar positions $(r_p,\th_p)$ but 
with opposite radial and polar velocities, $u^r,u^\th \rightarrow -u^r,-u^\th$.
Explicit calculations of the conservative and dissipative components of the
self-force have been made by identifying these reflection points
along restricted orbits (circular equatorial; eccentric equatorial; or
inclined spherical) \cite{Bara09,WarbBara11,Warb15,ThorWard17}.

\begin{figure}[bth]
\includegraphics[width=0.45\textwidth]{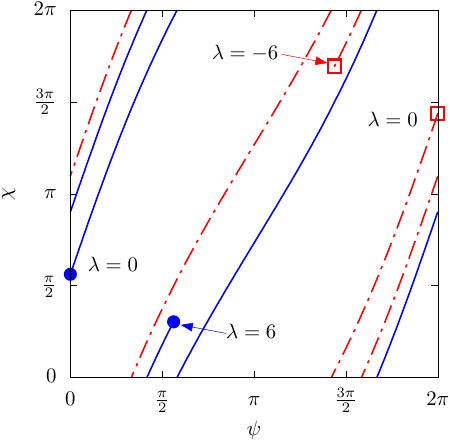}
\caption{Two orbits with the same orbital parameters
    $(p,e,\iota,a/M)=(5,0.6,1.04954,0.95)$ but different initial positions
mapped to the two-torus defined by the rotational coordinates $\psi$ and 
$\chi$.  The blue (solid) line traces an orbit that begins at Mino time 
$\la=0$ with initial position $(r_p,\th_p)=(r_\text{min},1.7409)$ and is 
terminated at $\la=6$.  This orbit follows from choosing $\la^{(r)}_0=0$ and 
$\la^{(\th)}_0=0.587813$ in Eqs.~\eqref{eqn:lambdar} and 
\eqref{eqn:lambdath}.  The red (dot-dashed) line follows an orbit with the 
reversed parameters, $\la^{(r)}_0=0$ and $\la^{(\th)}_0=-0.587813$, backward 
in time from $\la=0$ to $\la=-6$.  The points $\la=-6$ and $\la=6$ are 
example reflection points at which we can relate the advanced force 
$F^\text{adv}_\a$ to the retarded force $F^\text{ret}_\a$ using 
Eq.~\eqref{eqn:retAdvPsiChi}.} 
\label{fig:torus}
\end{figure}

For eccentric inclined orbits, these reflection points can be conveniently 
identified by mapping the particle's motion to a two-torus, as shown in 
Fig.~\ref{fig:torus}.  In this figure we cover the torus using the coordinates 
$\psi$ and $\chi$, related to the position in the polar ($r,\th$) plane by 
Eq.~\eqref{eqn:newpars}.  (Alternatively, some authors use the two angle 
variables $q_{r,\th}=\Upsilon_{r,\th}\la$ \cite{HindFlan08,Vand16,Vand17} to 
cover the torus.)  The polar motion winds and wraps in this region, either 
a finite number of times for a resonant orbit or an infinite number of times 
for a nonresonant orbit.  In the later case, the motion is ergodic and the 
motion will eventually pass all points arbitrarily closely.  All of the field 
and self-force information can be projected onto the domain spanned by 
$\psi,\chi\in[0,2\pi)$.

As an example, consider an orbit with geometric parameters
$(p,e,\iota,a/M)=(5,0.6,1.04954,0.95)$ and initial position
$(r_p,\th_p)=(r_\text{min},1.7409)$ set by taking $\la^{(r)}_0 = 0$ and 
$\la^{(\th)}_0=0.587813$, where $\lambda$ is measured in units of $M^{-1}$.
The path
of this orbit on the two-torus from $\la=0$ to $\la=6$ is traced out by the
blue (solid) line in Fig.~\ref{fig:torus}.  For any point on this curve, its
reflection point is identified by reflecting through the center of the 
plane at $\psi=\pi$ and $\chi=\pi$ (reflections can be made across any corner 
of the region equally well).  The result of reflecting the entire blue (solid) 
curve is the red (dot-dashed) curve.  This can be verified using
Eqs.~\eqref{eqn:newpars}-\eqref{eqn:Pth}.  Note that the red (dot-dashed) 
curve can also be described by an orbit moving backwards in time from $\la=0$ 
to $\la=-6$ with the same geometric parameters as the blue (solid) line, but 
with opposite offset: $\la^{(r)}_0=0$ but $\la^{(\th)}_0=-0.587813$.  This is 
in line with Eq.~(2.46) in \cite{Mino03}.

Therefore (up to a factor of $\pm 1$) the advanced force can be calculated by
reflecting the retarded force data on the torus.  Explicitly, the retarded and 
advanced forces are related by
\begin{equation} 
\label{eqn:retAdvPsiChi}
F^{\text{adv},l}_\a(\psi,\chi) 
= \eps_{(\a)} F^{\text{ret},l}_\a(2\pi-\psi,2\pi-\chi),
\end{equation}
where $\eps_{(\a)}=(-1,1,1,-1)$ and where there is no summation over $\alpha$.
Eq.~\eqref{eqn:retAdvPsiChi} can be extended to inclined spherical, eccentric
equatorial, and resonant orbits as well, though the motions within the torus 
are severely restricted for these special orbits.  

\begin{figure*}[!ht]
  \includegraphics[width=6.9in]{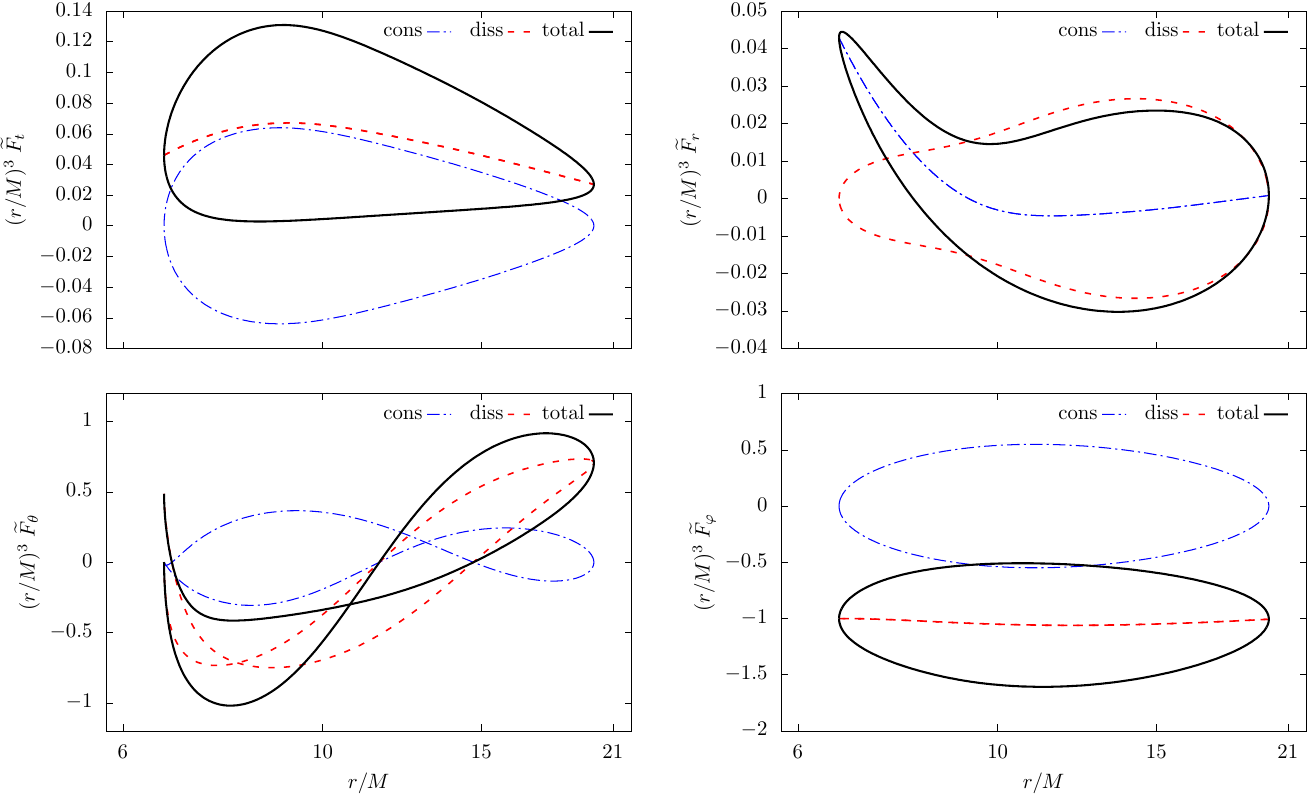}
  \caption{Components of the (dimensionless) scalar self-force for an inclined
  eccentric orbit in Schwarzschild spacetime.
  Note that we present our self-force results using the dimensionless
  quantities $\widetilde{F}_{t,r} \equiv (M^2/q^2) F_{t,r}$,
  and $\widetilde{F}_{\th,\vp} \equiv (M/q^2) F_{\th,\vp}$.
  The orbital parameters are given by $(p,e,\iota,a/M)=(10,0.5,\pi/5,0)$.
  The red (dashed) lines refer to the dissipative pieces of the self-force
  components, while the blue (dot-dashed) lines refer to the conservative
  pieces. The black (solid) lines represent the total values for each
  self-force component.
  $\widetilde{F}_t$ $\widetilde{F}_r$, $\widetilde{F}_\vp$ share the same
  periodicity as the particle's radial motion.
  Therefore, plotted as functions of $r$, these components form closed
  self-force ``loops." However $\widetilde{F}_\th$ does not close on itself in
  this eccentric inclined case, because $\widetilde{F}_\th$ also depends on
  the longitudinal position of the particle $\th_p$, which librates at a
  different frequency from the particle's radial position $r_p$
  ($\Omega_r\neq\Omega_\th$).} 
\label{fig:schwSSF}
\end{figure*}

\section{Results}
\label{sec:results}

Our results are broken down into three categories:
\begin{enumerate}[(a)]
  \item Eccentric inclined orbits in Schwarzschild spacetime;
  \item Highly eccentric equatorial orbits about a 
	rapidly rotating Kerr black hole, displaying quasinormal bursts;
  \item Eccentric inclined (generic) orbits in Kerr spacetime.
\end{enumerate}

\subsection{Schwarzschild eccentric inclined orbits}
\label{sec:schwOrbits}

\begin{figure*}[!t]
\includegraphics[width=6.7in]{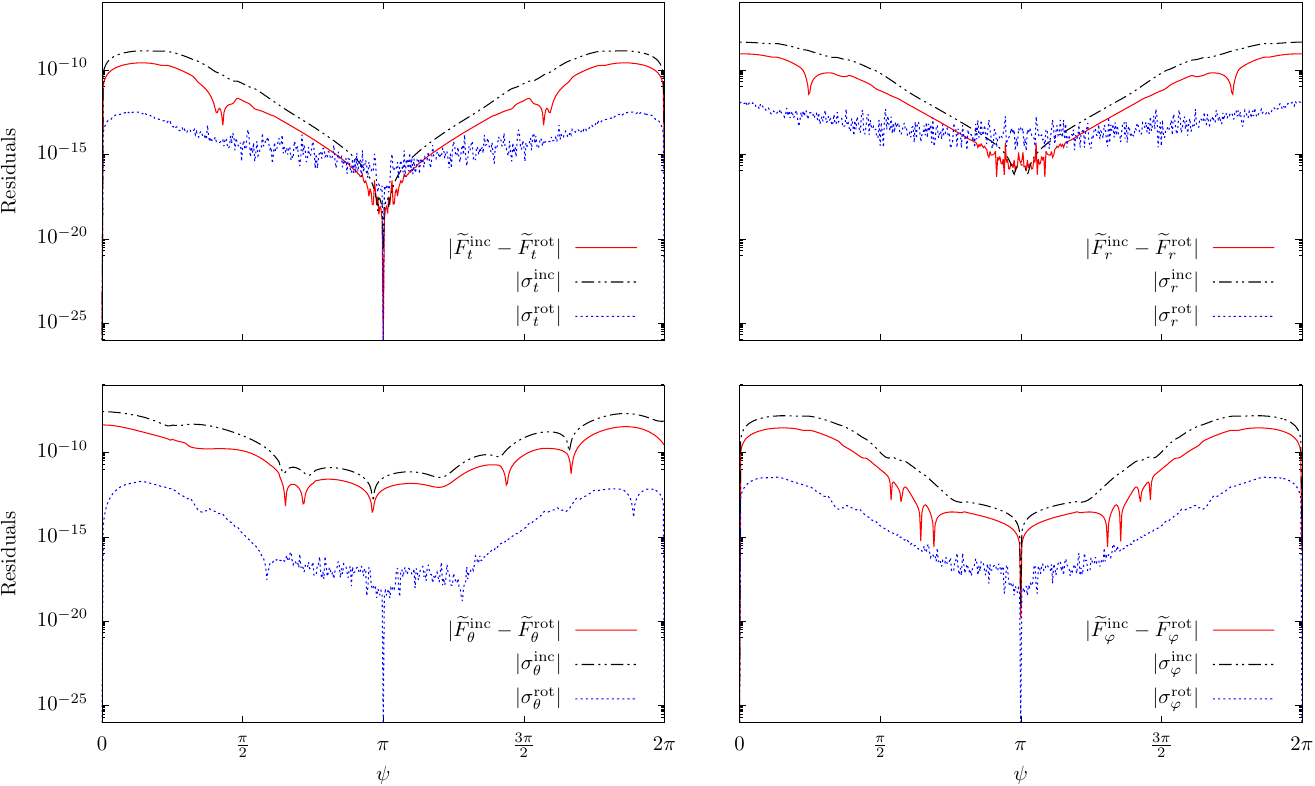}
\caption{Comparison of the scalar self-force calculated from an inclined 
orbit and a rotated equatorial orbit in Schwarzschild spacetime. The equatorial 
orbit is described by the orbital parameters $(p,e,\iota,a/M)=(10,0.5,0,0)$, 
while the inclined orbit is described by $(p,e,\iota,a/M)=(10,0.5,\pi/5,0)$. 
Red (solid) lines refer to the absolute residuals between the self-force 
calculated by rotating the results from an equatorial orbit 
$\widetilde{F}^{\text{rot}}_\a$ and the scalar self-force directly calculated 
from 
the inclined orbit $\widetilde{F}^{\text{inc}}_\a$.  The black (dot-dashed) and 
blue 
(dotted) lines refer, respectively, to the errors from calculating the 
self-force along an inclined orbit and an equatorial orbit.  The error for 
both the rotated equatorial orbit $\s_\a^{\text{rot}}$ and the error for the 
inclined orbit $\s_\a^{\text{inc}}$ are based on the estimated error from 
fitting the conservative 
component of the self-force, as outlined in Sec.~\ref{sec:modeSumReg}.} 
\label{fig:schwResiduals}
\end{figure*}

We first examine eccentric inclined orbits in the Schwarzschild limit 
($a = 0$).  These models serve as a strong validation of the SSF code, 
since all elements of the field and self-force calculation are required, yet 
they can be compared to much simpler-to-compute eccentric equatorial models 
(i.e., ones with vastly fewer computed modes).  The one-to-one correspondence 
results from spherical-symmetry of Schwarzschild spacetime, where two 
geodesics with the same eccentricities but different inclinations are related 
merely by a rotation.  

In spherically symmetric spacetimes, the self-force for an eccentric inclined 
orbit $F_\a$ can be compared to the force $F^{\text{rot}}_\a$ 
that is obtained through rotational transformation of the equatorial plane 
self-force $F^{\text{eq}}_\a$.  The transformation is
\begin{align}
F^{\text{rot}}_t &= F^{\text{eq}}_t, \label{eqn:ftEq} \\
F^{\text{rot}}_r &= F^{\text{eq}}_r, \label{eqn:frEq} \\
F^{\text{rot}}_\th &= \pm F^{\text{eq}}_\vp \; 
\sqrt{1 - \cos^2 \iota \csc^2\th_p}, \label{eqn:fthEq} \\
F^{\text{rot}}_\vp &= F^{\text{eq}}_\vp \cos\iota, \label{eqn:fphEq}
\end{align}
where $\pm$ depends on the sign of $u^\th$ ($+$ when $u^\th>0$).

The four SSF components for an orbit characterized by 
$(p,e,\iota,a/M)=(10,0.5,\pi/5,0)$ are plotted in Fig.~\ref{fig:schwSSF}.
For equatorial orbits, the self-force is a periodic function of $\psi$.  This
periodicity continues to be seen in Fig.~\ref{fig:schwSSF} for the $F_t$, 
$F_r$, and $F_\vp$ components in the inclined model as these self-force 
components ``loop'' back onto themselves as the particle librates from 
$r_\text{min}$ to $r_\text{max}$ and then back to $r_\text{min}$.  This 
periodicity is evident in examining Eqs.~\eqref{eqn:ftEq}, \eqref{eqn:frEq}, 
and \eqref{eqn:fphEq}.

The behavior of $F_\th$ is different.  When the orbit is rotated out of the 
equatorial plane, the $F^\text{eq}_\vp$ contribution is split between the 
rotated self-force components $F^{\text{rot}}_\vp$ and $F^{\text{rot}}_\th$.  
While $F^{\text{rot}}_\vp$ differs from $F^\text{eq}_\vp$ by a trigonometric 
factor, the projection of 
$F^\text{eq}_\vp$ onto the new inclined basis depends on the longitudinal 
position of the particle.  This causes $F^{\text{rot}}_\th$ to also 
depend upon $\th_p$ (see Eq. \eqref{eqn:fthEq}).  The small body librates at 
different frequencies in $r$ and $\th$, which demonstrates why the inclined 
force component $F_\th$ does not form a closed loop when plotted versus $r$.

These inclined SSF results can be compared in quantitative detail, again via 
Eqs.~\eqref{eqn:ftEq}-\eqref{eqn:fphEq}, to results computed from an equivalent 
equatorial orbit $(p,e,\iota,a/M)=(10,0.5,0,0)$.  We refer to the self-force
calculated directly using an inclined orbit as $F^\text{inc}_\a$, 
while the force computed by rotating the equatorial orbit self-force remains
being denoted by $F^\text{rot}_\a$.  The absolute residuals from 
comparing these orbits are plotted in Fig.~\ref{fig:schwResiduals}.  We also 
plot the errors $\sigma^\text{inc}_\a$ and $\sigma^\text{rot}_\a$ for both 
self-force calculations.  The primary source of error comes from fitting the 
conservative component of the self-force.  In Fig.~\ref{fig:schwResiduals} 
we see that the residual errors between the two calculations consistently fall 
below the errors that are estimated by our fitting procedure.  This provides 
additional confidence in the validity of our error estimation, which is 
outlined in Sec.~\ref{sec:modeSumReg}, and makes a strong case for having 
summed over all the required modes and correctly computed the regularization 
in the inclined model.

Additionally, we can compare specific numerical values of $F^\text{inc}_\a$ to
previously and independently computed equatorial results published in 
\cite{WarbBara11}, by again using Eqs.~\eqref{eqn:ftEq}-\eqref{eqn:fphEq} 
to transform the equatorial plane SSF.  We compare both the conservative and 
dissipative parts of the self-force in Table \ref{tab:schwCompare}.  The 
fractional errors between the independently computed conservative parts 
typically fall below the estimated errors in the conservative parts 
themselves that owe to the high-$l$ fitting procedure.  The dissipative part 
of our inclined SSF typically agrees with the transformed dissipative part 
from \cite{WarbBara11} to 6 or more decimal places.  

{\renewcommand{\arraystretch}{1.35}
\begin{table*}[!tbp]
\caption{A comparison between the scalar self-force data produced by our code 
for an eccentric inclined orbit $(p,e,\iota ,a/M)=(10,0.5,\pi/5,0)$ and 
equatorial scalar self-force results from Ref.~\cite{WarbBara11}.  We 
rotate the results of \cite{WarbBara11} using 
Eqs.~\eqref{eqn:ftEq}-\eqref{eqn:fphEq} to directly compare with our inclined 
values.  Conservative values include error estimates due to fitting the 
large-$l$ contribution as discussed in Sec.~\ref{sec:modeSumReg}.  Note that 
our fitting procedure, outlined in Sec.~\ref{sec:modeSumReg}, is partially 
motivated by but not equivalent to the fitting procedure in \cite{WarbBara11}. 
Numbers in parentheses describe the estimated error in the last reported 
digit, i.e. $1.44626(5) = 1.446(2) \pm 0.002$.  
Dissipative values are truncated based on the value of the last computed 
self-force $l$-mode $l_\text{max}$.}
\label{tab:schwCompare}
\begin{tabular*}{\textwidth}{c c @{\extracolsep{\fill}} S[table-format=2.7]
S[table-format=2.12] S[table-format=2.7]
S[table-format=2.12]}
\hline
\hline
& & \multicolumn{2}{c }{Conservative} & \multicolumn{2}{c }{Dissipative}
\\
\multicolumn{2}{c }{$\psi$} & \multicolumn{1}{c }{$0$} & \multicolumn{1}{c }{$\pi/2$} & \multicolumn{1}{c }{$0$} & \multicolumn{1}{c }{$\pi/2$}
\\
\hline
\multirow{2}{*}{$\widetilde{F}_t\times 10^4$} & This paper
& 0 & 0.5682633(2) & 1.5516959 & 0.657753715363
\\
& Rotated \cite{WarbBara11}
& 0 & 0.56825(3) & 1.5516962 & 0.65775426
\\
\multirow{2}{*}{$\widetilde{F}_r\times 10^4$} & This paper
& 1.44626(5) & -0.0306661(7) & 0 & 0.17666439973
\\
& Rotated \cite{WarbBara11}
& 1.446(2) & -0.0306717(7) & 0 & 0.17666437
\\
\multirow{2}{*}{$\widetilde{F}_\th\times 10^4$} & This paper
& 0 & -1.91200(1) & 0 & -3.726015695
\\
& Rotated \cite{WarbBara11}
& 0 & -1.9119(2) & 0 & -3.7260156
\\
\multirow{2}{*}{$\widetilde{F}_\vp\times 10^3$} & This paper
& 0 & -0.5392489(1) & -3.3771023 & -1.050859941917
\\
& Rotated \cite{WarbBara11} 
& 0 & -0.53923(6) & -3.3771019 & -1.0508599
\\
\hline
\hline
\\
\end{tabular*}
\end{table*}
}



\subsection{Highly eccentric orbit about a tapidly totating Kerr black hole 
and quasinormal bursts in the waveform}
\label{sec:eqOrbits}

Thornburg and Wardell \cite{ThorCapra14,ThorCapra16,ThorCapra17,ThorWard17} 
were the first to demonstrate that, for highly eccentric orbits
($e\gtrsim 0.7$) about rapidly rotating black holes ($a/M \gtrsim 0.8$),
interesting ``wiggles'' arise in the scalar self-force.  They further 
showed that these high frequency oscillations were attributable to excitation 
of a quasinormal mode (QNM), the least-damped $l=m=1$ mode, produced by 
periastron passage of the scalar-charged small body.  Thornburg and Wardell 
observed these excitations for a number of orbital configurations.  The most 
pronounced excitations were present in orbits with $e\geq 0.9$, though weak 
oscillations arise for the orbit $(p,e,\iota,a/M)=(8,0.8,0,0.8)$ (see 
Fig.~16 in \cite{ThorWard17}).

\begin{figure}[htb!]
\includegraphics[width=0.95\columnwidth]{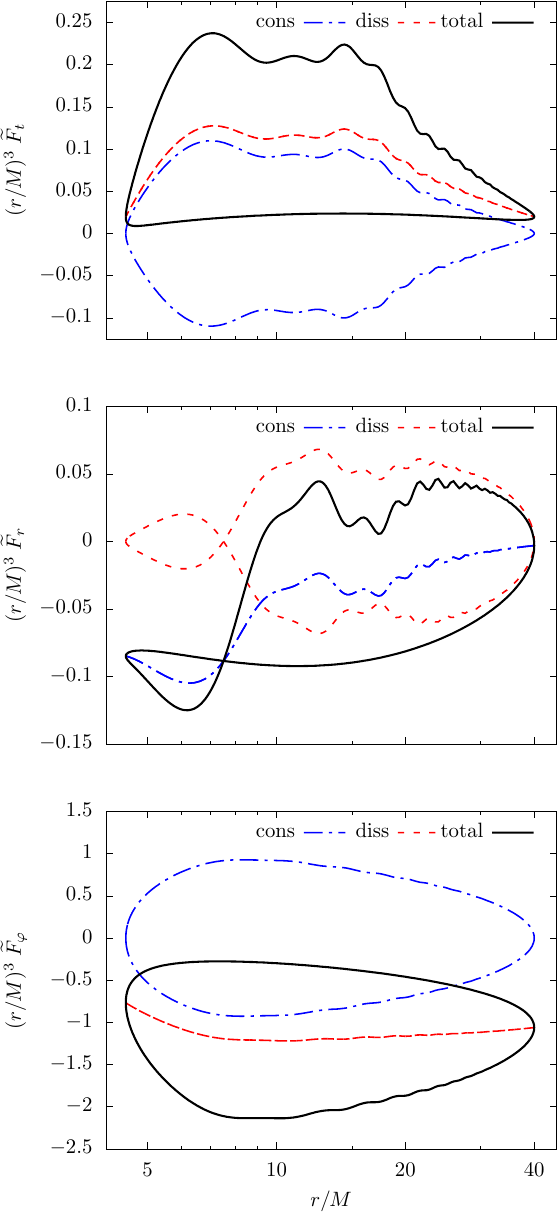}
\caption{The three nonzero components of the (dimensionless) scalar 
self-force for a particle orbiting in a Kerr background with orbital 
parameters $(p,e,\iota,a/M)=(8,0.8,0,0.99)$.  The red (dashed) lines refer 
to the dissipative pieces of the self-force components, while the blue 
(dot-dashed) lines refer to the conservative pieces.  The black (solid) lines 
represent the total values for each respective self-force component.} 
\label{fig:QNM}
\end{figure}

Thornburg and Wardell utilize a TD code, which can be well-suited for 
computing highly eccentric orbits.  However, TD codes involve solving 
partial differential equations and have potential numerical issues with 
initial value transients, boundary conditions, and source modeling.  Our code 
works in the frequency domain, where the numerical problem involves solving 
ordinary differential equations for large numbers of Fourier-harmonic modes.  
In general it is easier to attain higher accuracy with a FD code.  However, 
a countering factor is that the required number of modes and computational 
demand in a FD code grows exponentially at high eccentricities.  
Accordingly, we have so far restricted ourselves to orbits with $e\leq0.8$.  
On the positive side, a FD code only captures periodic behavior and is not 
subject to initial value transients.  Given the many differences between the 
two approaches, a comparison between results seemed desirable. 

Having said that, we have not made an exact comparison.  We have so far not 
tried to make a very time consuming calculation with $e=0.9$ to duplicate 
one of the results in \cite{ThorWard17}.  At the same time, rather than 
replicating the $e=0.8$ results of Thornburg and Wardell, with $a/M = 0.8$,
we decided to calculate the SSF and fluxes for the same orbital parameters 
but with a higher black hole spin: $(p,e,\iota,a/M)=(8,0.8,0,0.99)$.  The 
expectation was that we might see more pronounced ringing in the $e = 0.8$ 
orbit if the QNM damping is lessened with a higher $a/M$.

We also chose to model an orbit in the equatorial plane, which substantially 
offsets the computational cost of high eccentricity by restricting the 
mode spectrum $\o_{m0n}=m\O_\vp+n\O_r$ to be bi-periodic and not tri-periodic.
Additionally, higher-order regularization parameters are known for equatorial 
orbits \cite{HeffOtteWard14} and we were able to circumvent the fitting 
schemes discussed in Sec.~\ref{sec:modeSumReg} in this case, improving the 
convergence and reducing the estimated error.

Our FD SSF results for this model are plotted in Fig.~\ref{fig:QNM}.  The 
closed loops in the force components are split out into conservative part, 
dissipative part, and total.  We see the same oscillatory features in our 
self-force results as Thornburg and Wardell found, with the oscillations 
most prominent in the $t$ and $r$ self-force components.  After the point 
charge's periastron approach ($r\simeq 4.4M$), the ringing in the scalar 
field sweeps past the small body driving oscillations in the self-force, with 
the oscillations then decaying as the system approaches apastron.  As 
expected, by increasing the black hole spin, we observe a more persistent 
ringing compared to that seen in the Thornburg and Wardell $e=0.8$ model.

\subsubsection{Quasinormal bursts in the waveform and extracting multiple 
quasinormal modes}
\label{sec:fittingQNM}

As we mentioned in the Introduction, we decided to look at the waveform in 
this model to see if the excitations were present in the asymptotic field.  
While faint, there are indeed quasinormal bursts (QNBs) visible to most 
observers of the waveform.  The waveform itself, highlighted in the 
Introduction with Fig.~\ref{fig:waveform}, appears devoid of ringing at any 
of three observer angles: $(\th_\text{obs},\vp_\text{obs})=(\pi/2,0)$, 
$(\th_\text{obs},\vp_\text{obs})=(\pi/4,0)$, and 
$(\th_\text{obs},\vp_\text{obs})=(0,0)$.  However, high-pass filtering or 
emphasizing high frequencies, by taking two time derivatives of the waveform 
as shown in Fig.~\ref{fig:waveformDt2}, makes the bursts visible.  
Figure \ref{fig:waveformDt2} shows the second derivative measured by the 
observer at $(\th_\text{obs},\vp_\text{obs})=(\pi/2,0)$.  Similar excitation 
is visible to an observer at $(\th_\text{obs},\vp_\text{obs})=(\pi/4,0)$, but 
the QNBs are not present for an observer at position 
$(\th_\text{obs},\vp_\text{obs})=(0,0)$ (i.e., along the polar axis).  As 
we show below, this is consistent with the ringing being due to (prograde) 
axial $l=m$ perturbations of the field in the Kerr geometry.  

Rather than emphasizing high frequencies by taking time derivatives of the 
signal, one can instead apply a high-pass filter to attenuate the lower 
frequency ``background.''  We construct a high-pass Butterworth filter using 
\textit{Mathematica}'s \texttt{ButterworthFilterModel}, 
\texttt{ToDiscreteTimeModel}, and \texttt{RecurrenceFilter}.  We choose the 
filter's parameters by inspecting the power spectrum of the waveform.

\begin{figure}[ht!]
\includegraphics[width=0.95\columnwidth]{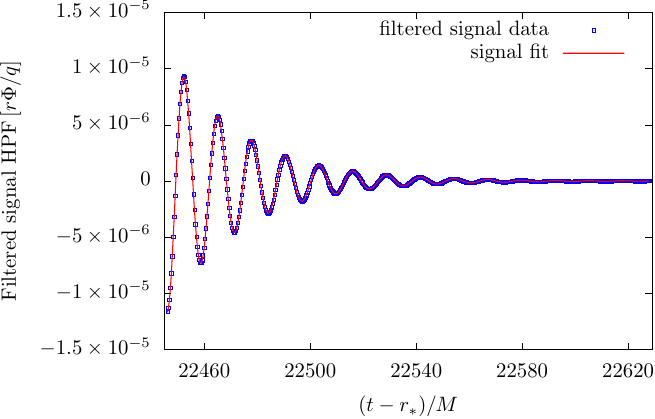}
\caption{Plot of a segment of the scalar field signal presented in 
Fig.~\ref{fig:waveform} after applying a high-pass filter (blue squares), 
along with a least-squares fit of the filtered signal (red line) to a 
model template.  The high-pass filter and fit were constructed as outlined 
in Sec.~\ref{sec:fittingQNM}.  The data are best fit by a decaying sinusoid 
with a complex frequency of $M\o=0.4933-0.0368i$.} 
\label{fig:waveformHP}
\end{figure}

After applying the high-pass filter and observing the presence of QNBs, we 
attempted to extract a complex frequency $\o = \o' + i\o''$ for the excitation 
by (1) selecting a time window during which the excitation dominates the 
filtered signal and (2) then performing a least-squares fit of a burst 
template to the filtered data, as demonstrated in Fig.~\ref{fig:waveformHP}.
The data was fitted to a real function of the form 
$A e^{+\o'' t}\sin\o'(t+t_0)$ using \textit{Mathematica}'s \texttt{FindFit}.  
Fitted complex frequencies have negative imaginary parts, consistent with 
damped bursts.  The data in Fig.~\ref{fig:waveformHP} was found to be best fit 
by the complex frequency $\o_\text{fit}=0.4937-0.0367i$ (in units with $M=1$; 
henceforth assumed in this section).

\begin{figure}[ht!]
\includegraphics[width=0.95\columnwidth]{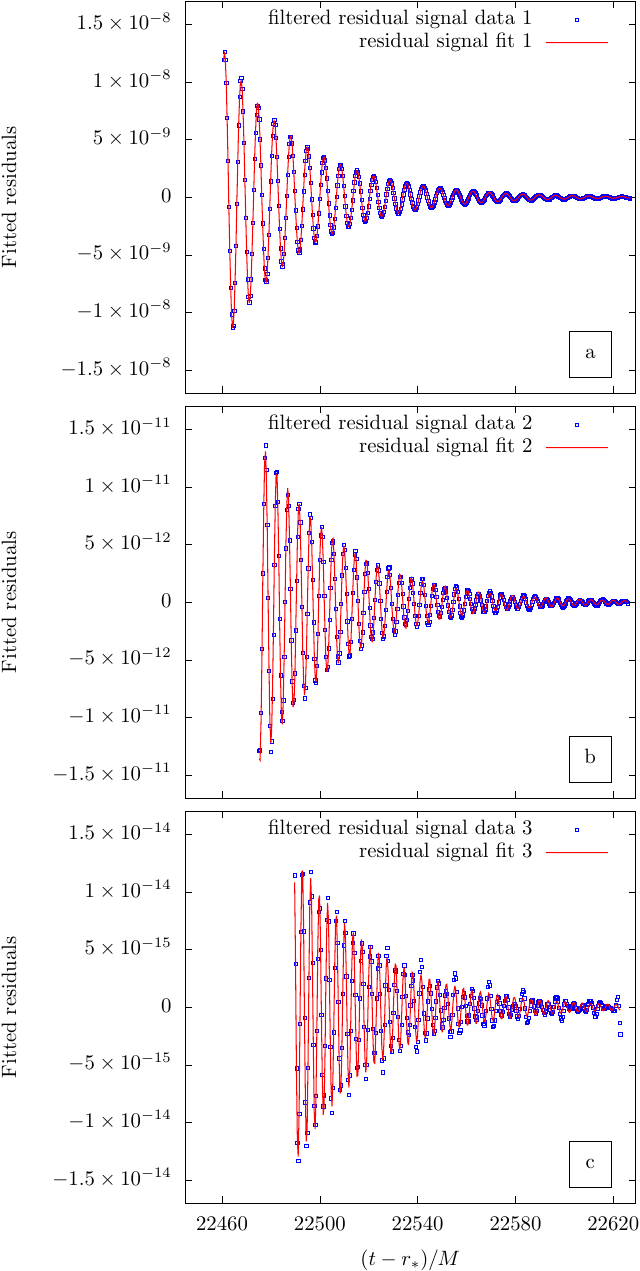}
\caption{Short window on the waveform showing successive sets of residuals 
(blue squares) after subtracting successively determined modes via fitting.  
Also shown are the least-squares determined fits of the residual signal data 
(red lines) at each stage in the subtraction.  The top plot (a) depicts the 
residual signal from subtracting the fit in Fig.~\ref{fig:waveformHP} from 
the waveform and high-pass filtering a second time.  The residuals in the 
top panel are then fit by a damped sinusoid with $M\o=0.9277-0.0314i$.  The 
middle panel (b) depicts the residuals after subtracting the first two QNMs 
and high-pass filtering.  The result is fit by a mode with 
$M\o=1.3682-0.0304i$.  The bottom panel (c) shows residuals after subtracting 
the first three determined QNMs and filtering, yielding a final mode with 
$M\o=1.8115-0.0304i$.  We found it necessary to slightly shift forward 
the time window after each fit.}
\label{fig:residualsFull} 
\end{figure}

We can compare this value to the spectrum of known QNM frequencies $\o_{plm}$ 
due to scalar perturbations of Kerr spacetimes published by Berti 
\cite{BertCardStar09}.  The QNMs depend on $a$ and are indexed by the 
spheroidal harmonic mode numbers $(l,m)$ and the overtone $p$, where $p=0$ 
refers to the least-damped or ``fundamental" overtone.  Assuming $M=1$ but 
without assuming a value for $a$, we find that the extracted complex frequency 
$\o_\text{fit}$ above most closely matches the QNM frequency 
$\o_{011}=0.4933-0.0368i$ for a spin of $a=0.9899$.  In other words, by 
assuming that this complex frequency should be represented by a QNM, the 
extracted frequency accurately recovers the spin of the primary black hole 
to three digits.  This result is consistent with those presented by Thornburg 
\cite{ThorCapra14,ThorCapra16,ThorCapra17}, who found that, across several 
orbital configurations and spin parameters, the QNM frequencies in his 
self-force data were best fit by the least-damped (smallest $|\o''|$) 
$l=m=1$ QNMs.

Surprisingly perhaps, our FD numerical results actually allow us to extract 
additional QNMs.  To do so, we obtain the residuals between the high-frequency 
signal and its fit in Fig.~\ref{fig:waveformHP} and apply the high-pass 
filter a second time to remove a remaining background (i.e., ``flat-fielding'' 
the signal).  We fit and obtain the complex frequency of a second damped 
oscillation.  By iterating this process, we managed to extract three 
additional QNM excitations in the filtered waveform.  These are shown in 
Fig.~\ref{fig:residualsFull}.  The numerical values of the frequencies of all 
extracted QNMs are presented in Table \ref{tab:QNM} and compared to the 
closest published QNMs for scalar perturbations of a Kerr spacetime with 
$a=0.99$.

However, we can instead try to remain agnostic to the black hole spin and mode 
numbers and compare the extracted frequencies to all known QNM frequencies 
across Berti's densely sampled set of Kerr spacetimes.  Consulting Table 
\ref{tab:QNM}, our second extracted frequency best fits a QNM in Berti's 
table with frequency $\o_{022}=0.9269-0.0314i$ for $a=0.9897$.  Our third
extracted frequency best fits one with $\o_{033}=1.3680-0.0304i$ for 
$a=0.9899$ and the fourth best fits Berti's mode $\o_{044}=1.8084-0.0304i$ for 
$a=0.9897$.  By simply looking for the best fit to 
known QNMs, we obtain multiple estimates of the black hole spin parameter.  
Multiple parameter estimates all yield values for the black hole spin that 
are surprisingly close to $a=0.99$ (with approximately three digits of 
agreement).  If QNBs can be observed in highly eccentric EMRIs, it may well 
be possible to get repeated snapshot determinations of the mass and spin of the 
primary black hole.  Furthermore, while the ``orbital parts'' of the EMRI 
waveform will evolve and move through the LISA passband, the frequencies of 
the QNB component of the waveform will remain invariant, as these depend upon 
the (essentially unchanging) primary mass and spin.

By reproducing Thornburg and Wardell's ``wiggles," we affirm that these are 
integral components of the SSF.  The finding of related QNBs in the scalar 
waveform suggests the strong likelihood that QNBs exist in the gravitational 
waveforms of (some) EMRIs.  A gauge invariant signal of this type, from 
repeatedly ``tickling'' the primary black hole, might have important 
observational consequences in sufficiently high signal-to-noise ratio EMRIs.

{\renewcommand{\arraystretch}{1.4}
\begin{table}
\caption{A comparison of the QNM frequencies extracted from filtering and 
fitting the waveform, as shown in Figs.~\ref{fig:waveformHP} and 
\ref{fig:residualsFull}, and the QNM frequencies calculated by Berti for scalar 
perturbations of Kerr spacetime with spin parameter $a/M=0.99$ 
\cite{BertCardStar09}. The value of $a$ is based on the spin parameter chosen 
for this highly eccentric SSF investigation.} \label{tab:QNM}
\begin{tabular*}{\columnwidth}{c @{\extracolsep{\fill}} c c c c c}
\hline
\hline
Figure & $p$ & $l$ & $m$ &	Extracted QNM & Known QNM \\
\hline
Fig.~\ref{fig:waveformHP} & 0 & 1 & 1 & $0.4933 - 0.0368i$ & 
$0.4934 - 0.0367i$ \\
Fig.~\ref{fig:residualsFull}(a) & 0 & 2 & 2 & $0.9277 - 0.0314i$ & 
$0.9280 - 0.0311i$ \\
Fig.~\ref{fig:residualsFull}(b) & 0 & 3 & 3 & $1.3682 - 0.0304i$ & 
$1.3686 - 0.0302i$ \\
Fig.~\ref{fig:residualsFull}(c) & 0 & 4 & 4 & $1.8115 - 0.0304i$ & 
$1.8111 - 0.0300i$ \\
\hline
\hline
\end{tabular*}
\end{table}
}

\subsection{Kerr inclined orbits}
\label{sec:genOrbits}

\begin{figure}[htb]
\includegraphics[width=0.95\columnwidth]{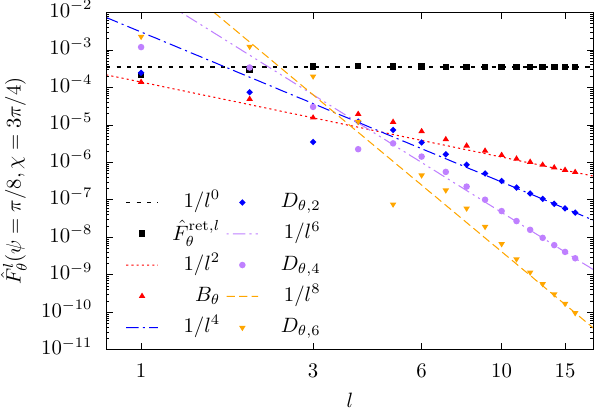}
\caption{Convergence of the (dimensionless) scalar self-force $l$-modes for 
an eccentric inclined orbit in Kerr spacetime.  Orbital parameters are 
taken to be $(p,e,\iota ,a/M)=(10,0.3,\pi/5,0.5)$.  The dashed and dotted 
lines depict the increasing rate of convergence for 
$\widetilde{F}_\th(\psi=\pi/8,\chi=3\pi/4)$ as additional regularization 
parameters are incorporated.  The black squares represent individual $l$-modes 
of the SSF prior to regularization, which diverge as expected.  The red 
triangles show the effect of subtracting the known analytic regularization 
parameters $A_\th$ and $B_\th$.  The blue diamonds include the next 
regularization parameter $D_{\th,2}$, estimated numerically 
(Sec.~\ref{sec:modeSumReg}).  The purple circles and the orange 
inverted triangles represent including additional numerically fitted 
regularization parameters.  Mode-sum convergence improves through 
inclusion of successively more regularization parameters.} 
\label{fig:regularizeGeneric}
\end{figure}

\subsubsection{Spherical inclined orbits}
\label{sec:compareSphOrbits}

We first examine inclined orbits in the Kerr background by calculating 
the SSF along spherical inclined orbits.  
Similar to other restricted orbits, spherical inclined orbits are biperiodic 
in their frequency spectrum, $\o_{mk0}=m\O_\vp+k\O_\th$, rather than 
tri-periodic like eccentric inclined orbits.  Additionally, while the number 
of summed radial-frequency modes in Eq.~\eqref{eqn:phiLM} rapidly grows with 
increasing eccentricities, the number of summed polar-frequency modes is not 
as dramatically affected by increasing the inclination.  Calculating the 
radial mode functions is also one of the primary computational bottlenecks 
of our code.  Altogether these factors significantly reduce computational 
costs, allowing us to compute the SSF along spherical 
orbits at large inclinations with high precision.

These orbits serve as a code test for us, since the SSF along spherical 
orbits was previously investigated by Warburton \cite{Warb15}.  We reproduced 
the results from \cite{Warb15} for the orbit with parameters 
$(p,e,\mathcal{L}_z/M,a/M)=(4,0,1,0.998)$.  To match the conventions of 
\cite{Warb15}, the orbit is parametrized by the $z$ component of angular 
momentum $\mathcal{L}_z$ instead of the inclination $\iota$.  The self-force 
data produced by our code are in good agreement with those of \cite{Warb15}. 
The conservative components agree to $\sim 4$ digits and dissipative 
components to 7 or more digits.  Comparative SSF values are provided in
Table \ref{tab:circCompare}.

{\renewcommand{\arraystretch}{1.35}
\begin{table*}[!tbp]
\caption{A comparison between the scalar self-force data produced by our code 
for a spherical inclined orbit $(p,e,\mathcal{L}_z/M,a/M)=(4,0,1,0.998)$ and 
the SSF results for the same orbit reported in Tables II and III of
\cite{Warb15}.  Conservative values include error estimates due to fitting the 
large-$l$ contribution as discussed in Sec.~\ref{sec:modeSumReg}. 
Numbers in parentheses describe the estimated error in the last reported 
digit, i.e. $-2.9793(5) = -2.9793 \pm 0.0005$. Dissipative values are truncated 
based on the value of the last computed dissipative self-force $l$-mode 
$l_\text{max}$.}
\label{tab:circCompare}
\begin{tabular*}{\textwidth}{c c @{\extracolsep{\fill}} S[table-format=2.14]
S[table-format=2.11] S[table-format=2.11]}
\hline
\hline
\multicolumn{2}{c }{$\psi$} & \multicolumn{1}{c }{$0$} & \multicolumn{1}{c }{$\pi/3$} & \multicolumn{1}{c }{$\pi/2$}
\\
\hline
\multirow{2}{*}{$\widetilde{F}_t^\text{cons}\times 10^4$} & This paper
& 0 & 1.077533(4) & 0
\\
& \cite{Warb15}
& 0 & 1.07740(5) & 0
\\
\multirow{2}{*}{$\widetilde{F}_t^\text{diss}\times 10^3$} & This paper
& 1.68377101827396 & 1.62358501378 & 1.66864142101
\\
& \cite{Warb15}
& 1.683771 & 1.623585 & 1.6686414
\\
\multirow{2}{*}{$\widetilde{F}_r^\text{cons}\times 10^4$} & This paper
& 4.0503727(9) & -3.901868(4) & -7.71977(2)
\\
& \cite{Warb15}
& 4.05036(4) & -3.90190(8) & -7.72001(4)
\\
\multirow{2}{*}{$\widetilde{F}_r^\text{diss}\times 10^4$} & This paper
& 0 & -1.28040714 & 0
\\
& \cite{Warb15}
& 0 & -1.2804071 & 0
\\
\multirow{2}{*}{$\widetilde{F}_\th^\text{cons}\times 10^3$} & This paper
& 3.5525351(2) & 2.25485(3) & 0
\\
& \cite{Warb15}
& 3.55243(9) & 2.25495(4) & 0
\\
\multirow{2}{*}{$\widetilde{F}_\th^\text{diss}\times 10^2$} & This paper
& 0 & -1.185212479 & -1.14620289587
\\
& \cite{Warb15}
& 0 & -1.1852125 & -1.1462029
\\
\multirow{2}{*}{$\widetilde{F}_\vp^\text{cons}\times 10^4$} & This paper
& 0 & -2.97984(2) & 0
\\
& \cite{Warb15} 
& 0 & -2.9793(5) & 0 
\\
\multirow{2}{*}{$\widetilde{F}_\vp^\text{diss}\times 10^3$} & This paper
& -4.96086992539137 & -7.2462959712 & -8.3045155780
\\
& \cite{Warb15}  
& -4.9608699 & -7.2462960 & -8.3045156
\\
\hline
\hline
\\
\end{tabular*}
\end{table*}
}

\subsubsection{Eccentric inclined orbits}
\label{sec:generic}

\begin{figure*}[htb]
\includegraphics[width=\textwidth]{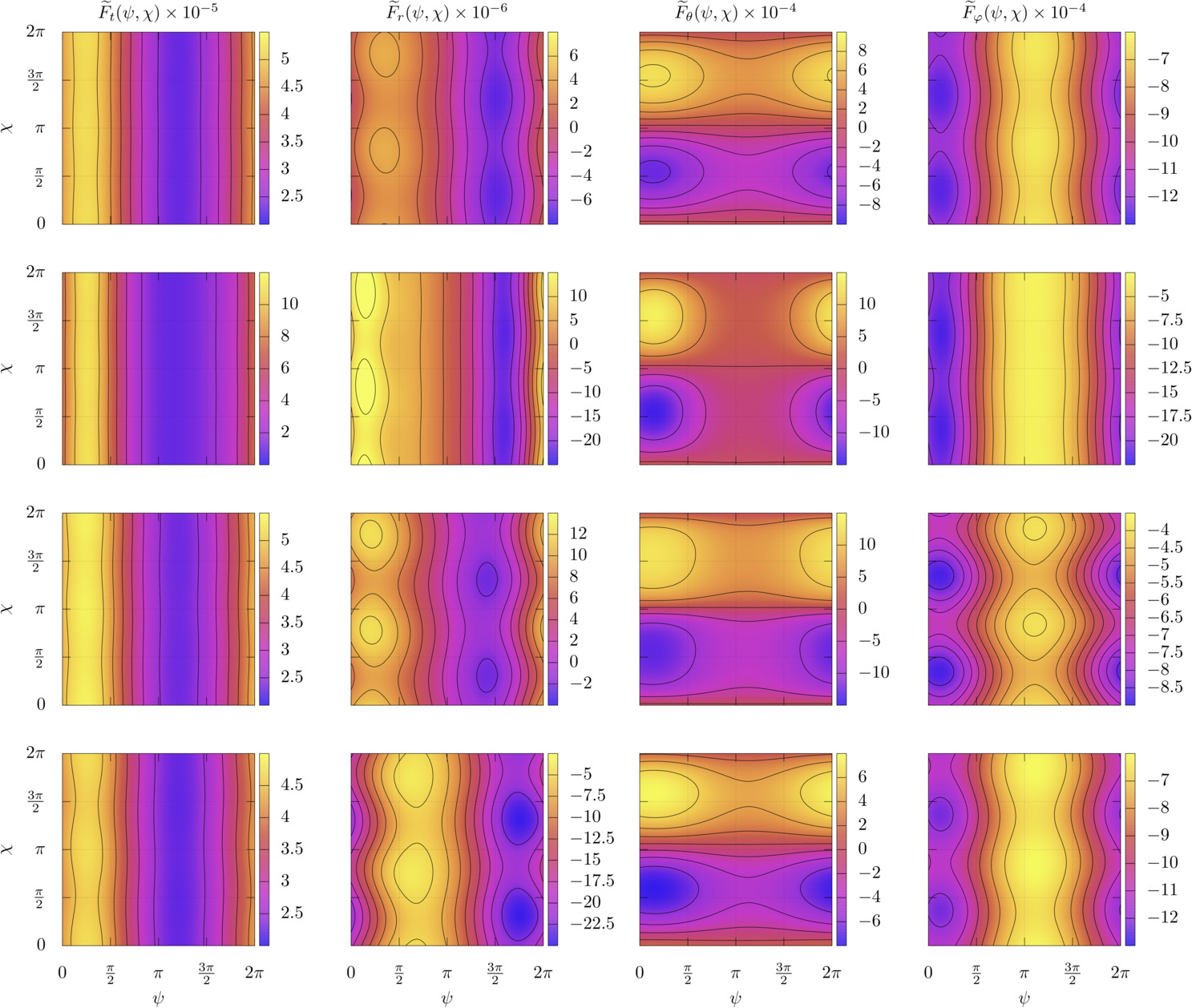}
\caption{The (dimensionless) scalar self-force components, 
$\widetilde{F}_\a(\psi,\chi)$, for the four orbits listed in Table 
\ref{tab:genOrbitParameters} is depicted through sampling on the torus.  
Each row of plots directly corresponds to the orbit in the same row of 
Table \ref{tab:genOrbitParameters}.  (The first, second, third, and fourth 
rows correspond to the orbits ``base," ``large e," ``large $\iota$," and 
``large a" respectively.)  The vertical axis is correlated with the 
$\th$-dependence of the self-force components, while the horizontal axis 
is related to the $r$-dependence.  Colors correspond to different values of 
the self-force, with the values denoted in the color bar to the right side of
each plot.  The self-force is constant along each contour line.  The tic 
labels in each colorbar correspond to the values of the contour lines.  
Therefore, in the top left plot, $\widetilde{F}_t=5\times 10^{-5}$ along 
the leftmost contour line.} 
\label{fig:generic}
\end{figure*}

The truly unique capability of our code is in being able to model the SSF on 
generic (bound) eccentric inclined orbits.  We investigate in this paper 
the SSF on four different orbits of this type, with their characteristic 
parameters specified in Table \ref{tab:genOrbitParameters}.  We refer to 
these orbits by their reference names: `base', `large $e$', `large $\iota$',
and `large $a$'.  We use the orbit $(p,e,\iota,a/M)=(10,0.1,\pi/5,0.5)$ as 
our fiducial case and then vary either the orbital eccentricity, the orbital 
inclination, or the black hole spin to get a sense of how the self-force 
depends on these orbital and spin parameters.  This also provides tests of 
our code's ability to probe more challenging regions of parameter space.  
The `large $e$' orbit is also used in Fig.~\ref{fig:regularizeGeneric} to 
demonstrate improved convergence of the mode-sum through incorporating 
additional numerically extracted regularization parameters.

{\renewcommand{\arraystretch}{1.4}
\begin{table}[bh]
\caption{Orbital parameters for generic orbits presented in 
Fig.~\ref{fig:generic}.} 
\label{tab:genOrbitParameters}
\begin{tabular*}{\columnwidth}{c @{\extracolsep{\fill}} c c c c}
\hline
\hline
    model & $p$ & $e$ & $\iota$ & $a/M$ \\
    \hline
    base & $10$ & $0.1$ & $\pi/5$ & $0.5$ \\
    large $e$ & $10$ & $0.3$ & $\pi/5$ & $0.5$ \\
    large $\iota$ & $10$ & $0.1$ & $\pi/3$ & $0.5$ \\
    large $a$ & $10$ & $0.1$ & $\pi/5$ & $0.9$ \\
\hline
\hline
\end{tabular*}
\end{table}
}

While in restricted cases the self-force can be periodic, for generic orbits
the self-force is instead biperiodic.  As such, it is less practical to
plot the self-force as a function of time or radial position as in
Figs.~\ref{fig:schwSSF} and \ref{fig:QNM}.  Instead, as long as the orbit is 
not resonant in $r$ and $\th$ motion, we can map the self-force as contour 
levels on the torus spanned by the coordinates $\psi$ and $\chi$, similar to 
the use of the torus in the discussion surrounding Fig.~\ref{fig:torus} of 
Sec.~\ref{sec:consDispSSF}.  The ergodic nature of the particle's motion 
implies that the SSF is a smooth continuous field over $\psi$ and $\chi$, 
with any given point eventually sampled by the motion (see also 
\cite{Vand17}).  This representation of the SSF for the generic (nonresonant) 
orbits listed in Table \ref{tab:genOrbitParameters} is shown in 
Fig.~\ref{fig:generic}.  (In these plots we use $\psi$ and $\chi$ as 
coordinates rather than angle variables $q_{r,\th} = \Upsilon_{r,\th}\la$ 
as found in \cite{Vand17}.)

For the orbits presented in Fig.~\ref{fig:generic}, the largest variations in 
the scalar self-force occur in the radial direction, with the exception of the
$F_\th$ component.  Consequently, despite the low eccentricities considered,
$F_t$, $F_r$, and $F_\vp$ are most dependent on $\psi$, i.e. the radial motion 
of the small body.  We also see that the maxima and minima of each self-force 
component are shifted away from the turning points of the particle's motion 
$(\psi=0,\pi,2\pi; \chi=0,\pi,2\pi)$ and the particle's passage through the 
equatorial plane $(\chi=\pi/2,3\pi/2)$, as a result of conservative effects.  
These shifts are most easily recognized in $F_r$.

Taking the ``base" orbit as a fiducial result, we can also examine how the
self-force changes as we vary the orbital parameters $e$ and $\iota$ or the
spin parameter $a$.  For the `high $e$' orbit, we increase the
eccentricity from $e=0.1$ to $e=0.3$.  We see that the radial dependence of 
the self-force becomes further accentuated, due to the orbit's increased 
eccentricity.  Additionally, the maximum magnitude of the scalar self-force 
increases in every self-force component, most likely due to the particle's 
smaller pericentric distance at the higher eccentricity.

For the high $\iota$ orbit, we increase the inclination from $\iota=\pi/5$ 
to $\iota=\pi/3$.  The dependence of the scalar self-force on the particle's
polar ($\chi$) motion becomes more pronounced, as the particle sweeps out a
larger region above and below the equatorial plane.  Additionally, the radial
component of the scalar self-force shifts to become predominantly positive.  
A similar behavior is seen for inclined spherical orbits, where the average 
value of $F_r$ grows monotonically with inclination, as it ranges from 
$\iota=0$ to $\iota=\pi$ \cite{Warb15}.  (Retrograde orbits are parametrized 
with $a<0$ in our code.)

{\renewcommand{\arraystretch}{1.4}
\begin{table*}[hbt!]
\caption{Energy and angular momentum fluxes for various orbits, along with 
their comparisons to the local work and torque done by the scalar self-force on 
the particle.  The plus signs in columns six and eight are due to the negative 
signs in Eqs.~\eqref{eqn:workEnBalance} and \eqref{eqn:torqueLzBalance}.  Flux 
expressions are truncated two digits prior to the order of the last calculated 
scalar self-force $l$-mode, 
$l_\text{max}$.  If the energy flux for $l_\text{max}$ is on the order of 
$10^{-14}$, then the flux is reported to an accuracy of $10^{-12}$.  The fluxes 
typically agree with the local work and angular momentum beyond the level of 
reported accuracy (the relative errors are greater than the reported accuracy 
of the results).  Note that the inclination for the last orbit corresponds to 
an angular momentum value of $\mathcal{L}_z/M=1$.} 
\label{tab:fluxBalance}
\begin{tabular*}{\textwidth}{ @{\extracolsep{\fill}} c c c c 
S[table-format=4.14] S[table-format=3.3] S[table-format=4.14] 
S[table-format=3.3]}
    \hline
    \hline
    $p$ & $e$ & $\iota$ & $a/M$ & \multicolumn{1}{c}{$\langle \dot{E}\rangle\; 
\times M^2/q^2$} & \multicolumn{1}{c}{$|1+\langle \dot{E} 
\rangle/\mathcal{W}|$} & \multicolumn{1}{c}{$\langle \dot{L}_z \rangle \; 
\times M/q^2$} & \multicolumn{1}{c}{$|1+\langle 
\dot{L}_z\rangle/\mathcal{T}|$} \\
    \hline
$10$ & $0.5$ & $\pi/5$ & $0$ & \num{3.32933297d-5} & \num{1d-11} & 
\num{6.34648550d-4} & \num{3d-10} \\
$10$ & $0.5$ & $0$ & $0$ & \num{3.32933297d-5} & \num{3d-11} & 
\num{7.84468749d-4} & \num{2d-11} \\
$10$ & $0.3$ & $\pi/5$ & $0.5$ & \num{2.9610263d-5} & \num{9d-14}  & 
\num{6.9840212d-4} & \num{4d-14}  \\
$10$ & $0.1$ & $\pi/3$ & $0.5$ & \num{2.994475370d-5} & \num{0d-11} & 
\num{4.93896206d-4} & \num{0d-12}  \\
$10$ & $0.1$ & $\pi/5$ & $0.9$ & \num{2.745901231d-5} & \num{7d-12} & 
\num{7.281232718d-4} & \num{0d-11}  \\
$10$ & $0.1$ & $\pi/5$ & $0.5$ & \num{2.917529922d-5} & \num{5d-14} & 
\num{7.56756034d-4} & \num{6d-15}  \\
$8$ & $0.8$ & $0$ & $0.99$ & \num{3.1363d-5} & \num{7d-8}  & 
\num{4.2122d-4} & \num{7d-9}  \\
$4$ & $0$ & $\sim 1.22$ & $0.998$ & \num{9.6423399d-4} & \num{7d-10}  & 
\num{3.7876524d-3} & \num{8d-10}  \\
    \hline
    \hline
\end{tabular*}
\end{table*}
}

For the high $a$ orbit, we increase the black hole spin parameter from 
$a/M=0.5$ to $a/M=0.9$.  We observe a stronger dependence of the scalar 
self-force on the polar position of the particle when $a$ is increased.  
Also, the radial component of the SSF becomes attractive ($F_r<0$) along the 
entire orbit in this case.  This is consistent with previous work on circular 
equatorial orbits, where $F_r$ decreases with increasing $a$ \cite{WarbBara10}.

\subsubsection{Flux balance}
\label{sec:fluxBalance}

As a final self-consistency check, we analyze the balance between the 
asymptotic fluxes with the local dissipative self-force effects
\cite{Mino03,Mino05,Mino05b,SagoETC06,WarbBara11}.  The average work done on 
the particle by the SSF should be balanced by the rate of radiative energy 
loss.  Likewise there should be a balance between the local torque on the 
particle due to the SSF and the angular momentum radiated away by the scalar 
field.  The average local work and torque are given, respectively, by
\begin{align}
\mathcal{W} & = - \lim_{T\rightarrow\infty} \frac{1}{T} \int_0^{T} 
\frac{F^\text{diss}_t}{u^t} dt, 
\label{eqn:localWork} \\
\mathcal{T} & = \lim_{T\rightarrow\infty} \frac{1}{T} \int_0^{T} 
\frac{F^\text{diss}_\vp}{u^t} dt. 
\label{eqn:localTorque}
\end{align}
In practice, periodicity (or bi-periodicity) can be leveraged to compute
Eqs.~\eqref{eqn:localWork} and \eqref{eqn:localTorque} with finite integrals
over time or finite integrals over the two-torus.  Note that only the 
dissipative component of the self-force contributes because both 
$F^\text{cons}_t$ and $F^\text{cons}_\vp$ are time-antisymmetric.  Therefore 
the conservative pieces cancel when averaging.

The asymptotic energy and angular momentum fluxes can be calculated by 
analyzing the scalar field at $r\simeq\infty$ and $r\simeq r_+$
\begin{align}
\langle \dot{E} \rangle &= \frac{1}{4\pi} \sum_{lmkn} \o_{mkn} 
\left( \g_{mkn} |C^-_{lmkn}|^2 + \o_{mkn} |C^+_{lmkn}|^2 \right), \\
\langle\dot{L}_z \rangle &= \frac{1}{4\pi} \sum_{lmkn} m 
\left(\g_{mkn} |C^-_{lmkn}|^2 + \o_{mkn} |C^+_{lmkn}|^2 \right),
\end{align}
where $E=\mu \mathcal{E}$, $L_z=\mu \mathcal{L}_z$, an overdot represents a
time derivative, and $\langle \rangle$ denotes a time ($t$) average.  Also 
recall that $\g_{mkn}\equiv\o_{mkn}-ma/2 M r_+$.  The flux balance formulas 
then take the form
\begin{align}
&\langle \dot{E} \rangle = -\mathcal{W}, \label{eqn:workEnBalance} \\
&\langle \dot{L}_z \rangle = -\mathcal{T}. \label{eqn:torqueLzBalance}
\end{align}
The fluxes and self-force are calculated independently from one another.
Consequently, comparing our scalar self-force results with flux calculations
provides a self-consistency check for our code.  Flux balance comparisons are
included in Table \ref{tab:fluxBalance}.

\section{Summary}
\label{sec:summary}

We considered a point scalar charge following generic bound geodesics
in Kerr spacetime and have calculated the scalar self-force acting on it, as 
a model for the gravitational self-force problem.  A \textit{Mathematica} 
code was designed to perform these calculations in the frequency domain with 
arbitrary numerical precision (we are currently developing C code 
to accomplish the same goals with increased computational efficiency).  Our 
numerical strategy includes novel features such as fast spectral source 
integration techniques that reduce expensive 2D source integrals to 
successive 1D Fourier sums.  We apply the same techniques to integrate the 
geodesic equations of motion.  The source calculation in the scalar case is 
sped up by orders of magnitude and argues for a thorough investigation of 
whether in the gravitational case the Teukolsky equation source can be 
similarly arranged to allow faster numerical integration. 

The accuracy of our code was validated by comparing to prior calculations 
of and existing results on the SSF, such as for (1) eccentric equatorial 
orbits, (2) inclined spherical orbits, and (3) self-comparison between 
inclined eccentric Schwarzschild and equatorial eccentric Schwarzschild.  
In all cases we verify that we calculate the scalar field and self-force 
with accuracy.  

In the process of computing the SSF on highly eccentric ($e=0.8$) equatorial 
orbits about a rapidly rotating ($a/M = 0.99$) Kerr primary, we verified a 
result of Thornburg and Wardell \cite{ThorWard17}--the existence of 
``wiggles" in the self-force due to quasinormal-mode excitation of the 
primary following periastron passage.  Their calculations were done with a 
time domain code while ours were done in the frequency domain.  Given 
substantial differences in the methods, it is heartening to see the result 
confirmed.

Intriguingly, we further searched for and observed quasinormal bursts 
(shortened to QNBs earlier in the paper) in the asymptotic waveform.  (This 
finding became a central highlight of the paper even though we have so far 
only computed it on equatorial orbits.)  We found that the QNBs are a 
superposition of not just the least-damped $l=m=1$ QNM (as \cite{ThorWard17} 
had already discovered) but of the least-damped $l=m=2,3,4$ QNMs as well.
While our calculations are of the scalar model problem, these QNBs are 
likely present in the gravitational waveform as well, which would provide a 
gauge-invariant indicator of the effect.  If so, these faint repeated bursts 
offer a new opportunity in high signal-to-noise ratio EMRI observations to 
measure rotating black hole properties.  In effect, each high $e$, high $a$ 
EMRI waveform would have two components: a low frequency spectrum that evolves 
toward higher frequency as the inspiral (chirp) proceeds and a high frequency 
spectrum of superposed damped modes which remain fixed in frequency (though 
with evolving amplitudes and phases).  It awaits future work to decide how 
practical measurement of QNBs might be in LISA observations given expected 
ranges on EMRI event rates.

Our results also focused on four different inclined eccentric orbits, with 
parameters given in Table \ref{tab:genOrbitParameters}, which represents the 
novel elements of our method and code.  We displayed in Fig.~\ref{fig:generic} 
how the scalar self-force changes from one of these orbits to the next, by
varying inclination, eccentricity, and black hole spin.  Validations of the 
generic orbit SSF results included examining convergence rates of the 
conservative self-force and checking balance between local SSF work and 
torque done on the small body and asymptotic energy and angular momentum 
fluxes.  

In future work we intend to apply the generic SSF code to study resonant 
orbits, directly measuring the size of jumps in the waveform that can be 
expected as a result of transient resonances and how those jumps vary with 
phase of the orbit upon entering the resonance \cite{FlanHind10,Vand14}.  
We will also likely make a thorough survey of QNB strengths, including 
moving beyond equatorial orbits.  Part of this work may focus on strategies 
for processing EMRI waveforms, e.g., matching templates or co-adding 
waveform segments, to try to draw QNBs up out of the detector noise.

\acknowledgments

We thank Niels Warburton, Adrian Ottewill, Barry Wardell, Maarten van de
Meent, Marc Casals, and Scott Hughes for helpful discussions.  This work was
supported in part by NSF Grants No.~PHY-1506182 and No.~PHY-1806447 
and by the North
Carolina Space Grant Graduate Research Fellowship.  C.R.E.~acknowledges
support from the Bahnson Fund at the University of North Carolina at Chapel 
Hill.
\\
\\
\noindent \textit{Note added in proof.}--After this paper was submitted a
related paper by Thornburg, Wardell, and van de Meent was submitted that
demonstrated the effect in the gravitational waveform \cite{ThorWardVand19}.
Also, the authors were made aware of an earlier paper by O'Sullivan and Hughes
\cite{OSulHugh16} where the quasinormal bursts were seen in the black hole
horizon shear response.

\begin{appendix}

\section{Regularizing the $\th$-component}
\label{app:thetaReg}

As mentioned in Sec. \ref{sec:modeSumReg}, we use a window function discovered 
by Warburton \cite{Warb15}

\be \label{eqn:windowFunc}
f(\th) = \frac{3 \sin^2\th_p\sin\th-\sin^3\th}{2\sin^3\th_p}.
\ee
This window function $f(\th)$ satisfies the necessary properties
$f\Phi \rightarrow \Phi$ and $\partial_\th\l f \Phi \r \rightarrow \partial_\th \Phi$ as $x^\mu \rightarrow x_p^\mu$,
ensuring that $F^{\text{ret}}_{\a\pm}$ is unaffected by the transformation
$\Phi\rightarrow f\, \Phi$. Additionally Warburton's window function
cleverly avoids wide bandwidth coupling
thanks to the compact relationship between $f\, \partial_\th Y_{jm}$
and $Y_{lm}$
\begin{align}
\label{eqn:fDY}
&f\, \partial_\th Y_{jm} = \b_{jm}^{(-3)} Y_{j-3,m} + \b_{jm}^{(-1)} Y_{j-1,m}
\\ &\qquad\qquad\qquad + \b_{jm}^{(+1)} Y_{j+1,m} +  \b_{jm}^{(+3)} Y_{j+3,m}. \notag
\end{align}
The coefficients $\b_{jm}^{(\pm i)}$ are defined as
\begin{gather}
\b_{lm}^{(\pm 1)} \equiv \l \frac{3\d_{lm}^{(\pm 1)}}{2\sin\th_p} - \frac{\zeta_{lm}^{(\pm 1)}}{2\sin^3\th_p}\r , \\
\b_{lm}^{(\pm 3)} \equiv \l\frac{\zeta_{lm}^{(\pm 3)}}{2\sin^3\th_p}\r ,
\end{gather}
where $\d_{lm}$ and $\zeta_{lm}$ are given in \cite{BaraSago10} as
\begin{gather}
\d_{lm}^{(+1)}=l C_{l+1,m}, \qquad \d_{lm}^{(-1)}=-(l+1)C_{lm}, \\
\zeta_{lm}^{(+3)}= -l C_{l+1,m}C_{l+2,m}C_{l+3,m}, \notag\\
\zeta_{lm}^{(-3)}= (l+1) C_{lm}C_{l-1,m}C_{l-2,m}, \notag\\
\zeta_{lm}^{(+1)}= C_{l+1,m}[l(1-C^2_{l+1,m}-C^2_{l+2,m})+(l+1)C^2_{lm}], \notag\\
\zeta_{lm}^{(-1)}= -C_{lm}[(l+1)(1-C^2_{l-1,m}-C^2_{lm})+lC^2_{l+1,m}], \notag\\
C_{lm}=\left[\frac{l^2-m^2}{(2l+1)(2l-1)}\right]^{1/2}. \notag
\end{gather}

Under these considerations, efficient calculation of $F^{\text{ret},l}_{\th\pm}$
follows from the replacement $\Phi\rightarrow f\, \Phi$
\begin{align}
&F^{\text{ret}}_{\th\pm} = q \lim_{x^\mu \rightarrow x^\mu_p} \sum_{j = 0}^{+\infty} \sum_{m=-j}^{j} \phi^\pm_{jm}(t,r) \, f(\th)\,  \partial_\th Y_{jm}(\th,\vp), \notag
\end{align}
\begin{align}
&\phantom{F^{\text{ret}}_{\th\pm}} = q \lim_{x^\mu \rightarrow x^\mu_p} \sum_{j = 0}^{+\infty} \sum_{m=-j}^{j} \phi^\pm_{jm}(t,r) \, \Big( \b_{jm}^{(-3)} Y_{j-3,m} \label{eqn:thFretMix}
\\&\q\;\; + \b_{jm}^{(-1)} Y_{j-1,m}+ \b_{jm}^{(+1)} Y_{j+1,m} + \b_{jm}^{(+3)} Y_{j+3,m} \Big). \notag
\end{align}
Refactoring Eq.~\eqref{eqn:thFretMix}, we recover Eq.~\eqref{eqn:fthExpansion}
\begin{align}
  &\psi^\pm_{lm}(t,r) = \b_{l+3,m}^{(-3)}\,\phi^\pm_{l+3,m}(t,r) + \b_{l+1,m}^{(-1)}\,\phi^\pm_{l+1,m}(t,r) \notag
  \\&\q\q + \b_{l-1,m}^{(+1)}\,\phi^\pm_{l-1,m}(t,r) + \b_{l-3,m}^{(+3)}\,\phi^\pm_{l-3,m}(t,r) .
\end{align}

\end{appendix}

\bibliography{specKerrPaper}

\end{document}